\documentclass[prc,	twocolumn,superscriptaddress,floatfix,	altaffillsymbol,nofootinbib]{revtex4-1}

\pdfoutput=1

\usepackage[fleqn]{amsmath} 
\usepackage{amssymb} 
\usepackage{array} 
\usepackage{bm} 
\usepackage{color} 
\usepackage{comment} 
\usepackage[T1]{fontenc} 
\usepackage{graphicx} 
\usepackage{lmodern} 
\usepackage{longtable} 
\usepackage{mhchem} 
\usepackage{microtype} 
\usepackage{scrextend} 
\usepackage{siunitx} 
\usepackage{subfigure} 
\usepackage{tabu} 
\usepackage{textcomp} 
\usepackage[referable]{threeparttablex} 
\usepackage{verbatim} 
\usepackage{xspace} 
\usepackage{hyperref} 

\DisableLigatures[f]{encoding = *, family = * } 

\begin{document}

\title{$\beta$ Decay of $^{132}$In and Spectroscopy of $^{132}$Sn and $^{131}$Sb with the GRIFFIN Spectrometer}

\date{\today}

\author{K.~Whitmore}
\affiliation{Department of Chemistry, Simon Fraser University, Burnaby, British Columbia, Canada V5A 1S6}
\author{C.~Andreoiu}
\affiliation{Department of Chemistry, Simon Fraser University, Burnaby, British Columbia, Canada V5A 1S6}
\author{F.H.~Garcia}
\affiliation{Department of Chemistry, Simon Fraser University, Burnaby, British Columbia, Canada V5A 1S6}
\author{K.~Ortner}
\affiliation{Department of Chemistry, Simon Fraser University, Burnaby, British Columbia, Canada V5A 1S6}
\author{J.D.~Holt}
\affiliation{TRIUMF, 4004 Wesbrook Mall, Vancouver, British Columbia, Canada V6T 2A3}
\affiliation{Department of Physics, McGill University, 3600 Rue University, Montr\'eal, Quebec, Canada H3A 2T8}
\author{T.~Miyagi}
\affiliation{TRIUMF, 4004 Wesbrook Mall, Vancouver, British Columbia, Canada V6T 2A3}
\author{G.C.~Ball}
\affiliation{TRIUMF, 4004 Wesbrook Mall, Vancouver, British Columbia, Canada V6T 2A3}
\author{N.~Bernier}
\altaffiliation{Present address: Department of Physics, University of the Western Cape, P/B X17, Bellville, ZA-7535 South Africa}
\affiliation{TRIUMF, 4004 Wesbrook Mall, Vancouver, British Columbia, Canada V6T 2A3}
\affiliation{Department of Physics and Astronomy, University of British Columbia, Vancouver, British Columbia, Canada V6T 1Z4}
\author{H.~Bidaman}
\affiliation{Department of Physics, University of Guelph, Guelph, Ontario, Canada N1G 2W1}
\author{V.~Bildstein}
\affiliation{Department of Physics, University of Guelph, Guelph, Ontario, Canada N1G 2W1}
\author{M.~Bowry}
\altaffiliation{Present address: School of Computing, Engineering and Physical Sciences, University of the West of Scotland, Paisley PA1 2BE, United Kingdom}
\affiliation{TRIUMF, 4004 Wesbrook Mall, Vancouver, British Columbia, Canada V6T 2A3}
\author{D.S.~Cross}
\affiliation{Department of Chemistry, Simon Fraser University, Burnaby, British Columbia, Canada V5A 1S6}
\author{M.R.~Dunlop}
\affiliation{Department of Physics, University of Guelph, Guelph, Ontario, Canada N1G 2W1}
\author{R.~Dunlop}
\affiliation{Department of Physics, University of Guelph, Guelph, Ontario, Canada N1G 2W1}
\author{A.B.~Garnsworthy}
\affiliation{TRIUMF, 4004 Wesbrook Mall, Vancouver, British Columbia, Canada V6T 2A3}
\author{P.E.~Garrett}
\affiliation{Department of Physics, University of Guelph, Guelph, Ontario, Canada N1G 2W1}
\author{J.~Henderson}
\altaffiliation{Present address: Lawrence Livermore National Laboratory, Livermore, California 94550, USA}
\affiliation{TRIUMF, 4004 Wesbrook Mall, Vancouver, British Columbia, Canada V6T 2A3}
\author{J.~Measures}
\affiliation{TRIUMF, 4004 Wesbrook Mall, Vancouver, British Columbia, Canada V6T 2A3}
\affiliation{Department of Physics, University of Surrey, Guildford GU2 7XH, United Kingdom}
\author{B.~Olaizola}
\affiliation{TRIUMF, 4004 Wesbrook Mall, Vancouver, British Columbia, Canada V6T 2A3}
\author{J.~Park}
\altaffiliation{Present address: Department of Physics, Lund University, 22100 Lund, Sweden}
\affiliation{TRIUMF, 4004 Wesbrook Mall, Vancouver, British Columbia, Canada V6T 2A3}
\affiliation{Department of Physics and Astronomy, University of British Columbia, Vancouver, British Columbia, Canada V6T 1Z4}
\author{C.M.~Petrache}
\affiliation{Universit\'{e} Paris-Saclay, CNRS/IN2P3, IJCLab, 91405, Orsay, France}
\author{J.L.~Pore}
\altaffiliation{Present address: Lawrence Berkeley National Laboratory, Berkeley, California 94720, USA}
\affiliation{Department of Chemistry, Simon Fraser University, Burnaby, British Columbia, Canada V5A 1S6}
\author{J.K.~Smith}
\altaffiliation{Present Address: Department of Physics, Pierce College, Puyallup, Washington 98374, USA}
\affiliation{TRIUMF, 4004 Wesbrook Mall, Vancouver, British Columbia, Canada V6T 2A3}
\author{D.~Southall}
\altaffiliation{Present address: Department of Physics, University of Chicago, Chicago, IL 60637, USA}
\affiliation{TRIUMF, 4004 Wesbrook Mall, Vancouver, British Columbia, Canada V6T 2A3}
\author{C.E.~Svensson}
\affiliation{Department of Physics, University of Guelph, Guelph, Ontario, Canada N1G 2W1}
\author{M.~Ticu}
\affiliation{Department of Chemistry, Simon Fraser University, Burnaby, British Columbia, Canada V5A 1S6}
\author{J.~Turko}
\affiliation{Department of Physics, University of Guelph, Guelph, Ontario, Canada N1G 2W1}
\author{T.~Zidar}
\affiliation{Department of Physics, University of Guelph, Guelph, Ontario, Canada N1G 2W1}

\begin{abstract}
Spectroscopy of doubly magic $^{132}_{50}$Sn$_{82}$ has been performed with the GRIFFIN spectrometer at TRIUMF-ISAC following the $\beta$ decay of $^{132}_{49}$In$_{83}$.
The analysis has allowed for the placement of a total of 70 transitions and 29 excited states in $^{132}$Sn.
Detailed spectroscopy has also been performed on $^{131}$Sb, resulting from the $\beta$ decay of $^{131}$Sn, produced from the $\beta$-delayed neutron decay of $^{132}$In.
Measurement of $\gamma$-rays in both $^{131}$Sn and $^{131}$Sb has led to the determination of the $\beta$-delayed neutron emission probability, $P_{n}$, from $^{132}$In.
This is the first time the $P_{n}$ has been measured for this nucleus using $\gamma$ spectroscopy, and the new value of 12.3(4)\% is consistent with the most recent $\beta-n$ counting experiment.
Additionally, $\gamma$-$\gamma$ angular correlations have been performed in $^{132}$Sn, supporting the spin assignments of several excited states.
Novel ab initio calculations are presented which describe several of the excited states, and these are compared to the experimental spectrum.
\end{abstract}

\maketitle

\section{\label{sec:introduction} Introduction}

The tin isotopes have been the subject of many experimental studies because of the magic number of protons with $Z=50$.
The known isotopes span more than a full shell closure and include both doubly magic $^{100}_{50}$Sn$_{50}$ and $^{132}_{50}$Sn$_{82}$.
Very little is known about $^{100}$Sn; no excited states have been identified, and only the mass, half-life and some $\beta$-decay properties have been studied~\cite{Chartier1996,Hinke2012,Lubos2019}.
In contrast, many properties of $^{132}$Sn are known.
With a first excited state above 4~MeV, it has been described as the nucleus with the strongest shell closure~\cite{Blomqvist1981,Fogelberg1994}, and it serves as an important benchmark for theoretical calculations.

Several $\beta$-decay experiments have identified excited states in $^{132}$Sn~\cite{Bjornstad1986,Fogelberg1994,Fogelberg1995}.
An early comprehensive $\beta$-decay study identified a total of 21 excited states and 44 transitions~\cite{Fogelberg1994,Fogelberg1995}.
From the same experiment, conversion electron spectroscopy revealed the $E1$ nature of the 311-keV transition, providing a $3^{-}$ assignment for the 4352-keV level.
A study of fission fragments from $^{248}\mathrm{Cm}$ identified a $9^{+}$ state at 5280~keV in delayed coincidence with the decay of the $8^{+}$ isomer, and also confirmed other yrast states up to 4942~keV~\cite{Bhattacharyya2001}.
The first $2^{+}$ and $3^{-}$ excited states have been further characterized through Coulomb excitation experiments, both at the Holifield Radioactive Ion Beam Facility, which measured the $B(E2; 0^{+}\rightarrow2^{+})$~\cite{Radford2004,Beene2004,Radford2005,Varner2005}, and at the High Intensity and Energy Isotope Separator On-Line (HIE-ISOLDE) facility, which additionally measured the $B(E3; 0^{+}\rightarrow3^{-})$~\cite{Rosiak2018}.
Another experiment at HIE-ISOLDE identified three new excited states from the $\beta$-delayed neutron ($\beta n$) decay of $^{133}\mathrm{In}$~\cite{Piersa2019}.
While the current manuscript was under review, the authors learned of a similar article discussing additional results from ISOLDE ~\cite{Benito2020} in which 68 new $\gamma$-ray transitions and 17 new levels were observed in $^{132}$Sn from both the $\beta$ decay of $^{132}$In and $\beta n$ decay of $^{133}$In.
Due to the timing of this publication, the results of this experiment are not discussed here.

Despite the availability of experimental data, there has been relatively little theoretical work characterizing the structure of $^{132}$Sn.
Several quasiparticle random-phase approximation (QRPA) calculations have been able to reproduce the measured $B(E2)$ and $B(E3)$ values~\cite{Terasaki2002,Ansari2005,Ansari2006,Yuksel2018}.
More recently, the first large-scale shell-model (LSSM) and Monte Carlo shell-model (MCSM) calculations have also reproduced the quadrupole and octupole transition strengths~\cite{Rosiak2018}.
Most of these investigations have also calculated the energies of the first $2^{+}$ and $3^{-}$ states, with generally good agreement with the measured values.
However, the theoretical calculations have been limited to these two low-lying excited states.
Additional work is therefore desired to examine the excited states beyond 5~MeV.

The region around $^{132}$Sn is especially important in rapid-neutron capture process ($r$-process) nucleosynthesis.
The solar system abundance peak at $A\sim130$ corresponds to nuclei along the $N=82$ closed shell~\cite{Mumpower2016}.
Sensitivity studies which determine the relative impact of nuclear properties on astrophysical element production reveal that $\beta$-decay properties, including half-lives and $\beta$-delayed neutron emission probabilities, $P_{n}$, exert a strong influence over the final abundances in this region.
The importance of determining neutron branching ratios, both for theoretical astrophysics and for heat production in reactors, has been highlighted by the International Atomic Energy Agency~\cite{Abriola2011}.

Several previous measurements of $\beta$-delayed neutron decay have been made by counting neutrons or $\beta$ particles.
More recently, however, several $\gamma$-spectroscopy experiments have provided updated and often more precise $P_{n}$ values for various nuclei~\cite{Jungclaus2016,Dunlop2019,Piersa2019}.
The first experiment to measure the $\beta$-delayed neutron decay from $^{132}$In found a $P_{n}$ of 4.2(9)\% using $\beta/n$ coincidences~\cite{Lund1980}.
Subsequent measurements using $\beta-n$ counting resulted in values of 6.8(14)\% and 10.7(33)\%~\cite{Reeder1986,Rudstam1993}.
An investigation of the $\beta n$ emission probability using the complimentary technique of $\gamma$ spectroscopy, which has a higher resolution and selectivity than neutron detection, is therefore valuable.

The present work describes one of the most detailed analyses of $^{132}$Sn, identifying 70 $\gamma$-ray transitions, along with 29 total excited states.
The decay of $^{132}$In has been characterized, including the half-life, $\beta$ feeding intensities, and the first measurement of $P_{n}$ using $\gamma$ rays.
For the first time, $\gamma$-$\gamma$ angular correlations in $^{132}$Sn have been performed, which support the spin assignments of several levels.
While ab initio theory has progressed to the light tin isotopes for both energies and $\beta$-decay rates~\cite{Morris2017,Gysbers2019}, it has been challenging to obtain converged calculations near $^{132}$Sn~\cite{Lascar2017,Manea2020}. 
Improved valence-space in-medium symmetry renormalization group (VS-IMSRG) calculations are presented, which are largely converged and are able to reproduce several low-lying levels in $^{132}$Sn.

\section{\label{sec:experiment} Experiment}

The experiment was conducted at the Isotope Separator and ACcelerator (ISAC) facility~\cite{Dilling2014} at TRIUMF.
A 9.8-$\mu$A beam of protons was accelerated to 480~MeV in the main cyclotron and impinged onto a $\mathrm{UC}_{x}$ target, inducing spallation and fission reactions.
The Ion Guide Laser Ion Source (IG-LIS)~\cite{Raeder2014} was used to suppress surface ionized elements.
A beam of $^{132}$In was selected by a high-resolution mass separator and sent to the experimental area in the ISAC-I hall at an energy of 28~keV.

The $^{132}$In beam was delivered to the Gamma-Ray Infrastructure For Fundamental Investigations of Nuclei (GRIFFIN)~\cite{Svensson2014,Garnsworthy2019}.
Signals were read out and processed by the GRIFFIN digital data acquisition system~\cite{Garnsworthy2017}.
GRIFFIN consists of 16 high-purity germanium clover detectors~\cite{Rizwan2016} arranged in a rhombicuboctahedral geometry.
The clovers were placed at a distance of 11~cm from the beam implant position, leading to an absolute efficiency of about 14\% at 1~MeV~\cite{Garnsworthy2019}.
The high efficiency and energy resolution of GRIFFIN facilitated the observation of $\gamma$-$\gamma$ coincidences used to build the level scheme in $^{132}$Sn.
In order to select $\gamma$ rays originating from the decay of $^{132}$In, the SCintillating Electron Positron Tagging ARray (SCEPTAR)~\cite{Garnsworthy2019} was used to detect electrons emitted through $\beta^{-}$ decay.
SCEPTAR was located inside the vacuum chamber and consists of 20 plastic scintillators and covers approximately 80\% of the solid angle around the beam implantation spot.
Because of the large $Q$-value of the $^{132}$In $\beta^{-}$ decay ($Q_{g.s.}=14.140(60)$~MeV~\cite{Wang2017}), a 2-cm thick Delrin shield was placed around the vacuum chamber to prevent high-energy electrons from reaching the GRIFFIN detectors.

The beam was implanted onto a mylar tape system at the center of GRIFFIN and SCEPTAR.
The tape was periodically moved to remove background events.
The $(7^{-})$ ground state of $^{132}$In has a half-life of 200(2)~ms~\cite{Singh132In}, and the activity on the tape saturated quickly.
Tape cycles were chosen to maximize the implantation time and total decays of $^{132}$In while reducing the activity from the subsequent decay of $^{132}$Sn ($T_{1/2}$=39.7(8)~s~\cite{Khazov2005}) as well as build-up of $^{132}\mathrm{Cs}$ contamination ($T_{1/2}$=6.480(6)~d~\cite{Khazov2005}).
A typical cycle consisted of tape moving and background measurement for 2~s, beam implantation for 30~s, and 5~s of beam decay.
During the background and decay portions of the cycle, the $^{132}$In beam was deflected away from the GRIFFIN beamline by an electrostatic kicker.
After cycling, the tape was moved into a lead-shielded box.
During this experiment, $^{132}$In was delivered at an average beam intensity of 70~pps over a period of 63~hours, providing a total of $5\times10^{7}$ $\beta$-gated $\gamma$-$\gamma$ coincidences.

Because of the short half-life of $^{132}$In, only data taken during the beam implantation cycle was used for the spectroscopy.
Contamination from $^{132}\mathrm{Cs}$ was minimal, and was almost completely eliminated by placing $\gamma$-$\gamma$ coincidences on known transitions in $^{132}$Sn.
Several strong peaks in the $\beta$-decay granddaughter $^{132}\mathrm{Sb}$ were observed, but these could also be eliminated from the analysis in the same way.

The relative efficiency of GRIFFIN was determined for the energy region up to 3.2~MeV using standard sources of $^{133}$Ba, $^{152}$Eu, $^{60}$Co, and $^{56}$Co.
Corrections to $\gamma$-ray intensities in the sources were made in order to account for real-coincidence summing.
These were made by building a 180\textdegree\ $\gamma$-$\gamma$ coincidence matrix as described in Ref.~\cite{Garnsworthy2019}.

In $^{132}$Sn, there are three transitions above 4~MeV, where it was not possible to directly determine the efficiency using available sources.
However, due to $\beta$-decay selection rules the $2^{+}$ state at 4.041~MeV is not directly populated by the $\beta$ decay of the $(7^{-})$ ground state in $^{132}$In.
Therefore, the relative efficiency above 4~MeV was tuned so that the intensity of the 4.041~MeV $2^{+}\rightarrow0^{+}$ transition matched the intensity of the 311- and 375-keV transitions which directly feed the $2^{+}$ state.
The intensities of these two low-energy transitions included the theoretical conversion coefficients calculated with the BrIcc code~\cite{Kibedi2008} with the frozen orbital approximation.
The relative efficiency curve determined from the sources was extrapolated out to 4.416~MeV, the highest-energy transition in the present analysis.
Comparing the extrapolated efficiency curve with the tuned efficiency at 4.041~MeV gave a scale factor which was applied to the extrapolated efficiencies at 4.352 and 4.416~MeV.

Data from GRIFFIN was collected in addback mode, where hits from different germanium crystals within the same clover were combined into a single event.
This method increased the photopeak signal while decreasing the Compton background.
The time window for building addback coincidences was 300~ns.
Summing corrections were also performed on the intensities of $^{132}$Sn $\gamma$ rays using 180\textdegree\ coincidences.
For weaker peaks which were only visible in gated spectra, summing corrections required more care.
In these cases, corrections were made by requiring a $\gamma$-$\gamma$-$\gamma$ triple coincidence condition in which the first hit was the strong $\gamma$ ray used to make the original gate.
A second requirement that the remaining two $\gamma$ rays be observed in detectors 180\textdegree\ apart provides a comparable analysis to the $\gamma$-$\gamma$ coincidence method described in Ref.~\cite{Garnsworthy2019}.
Signals from each crystal were linearly gain-matched using two strong peaks in $^{132}$Sn (375 and 4041~keV).
Other relatively strong peaks were used to ensure that energies were well aligned among all the crystals.

\begin{figure}[t]
\includegraphics[width=\columnwidth]{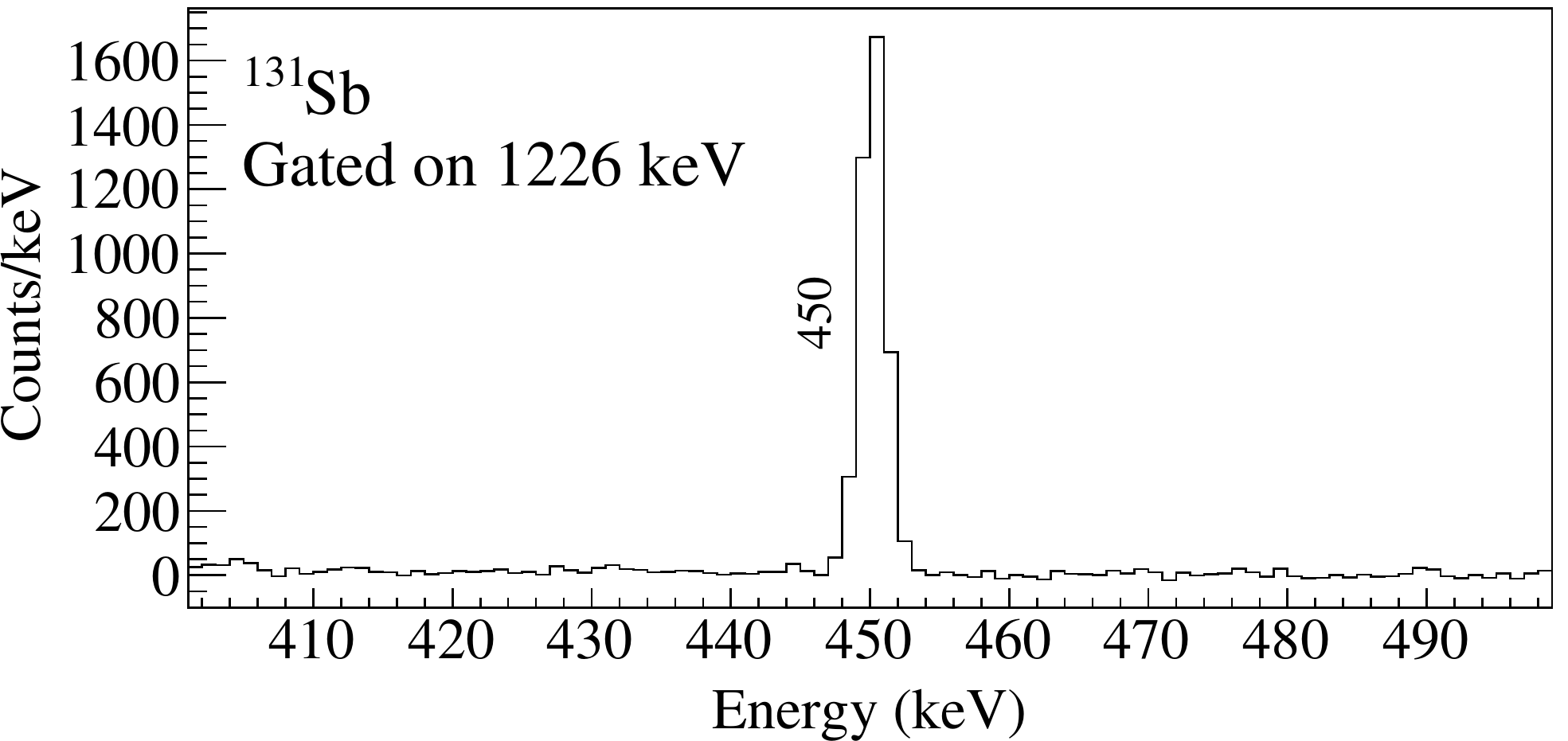}
\caption{\label{fig:131Sbspec}
A sample of the $\gamma$ spectrum gated on the 1226-keV $(11/2^{+})\rightarrow(7/2^{+})$ transition in $^{131}$Sb, clearly demonstrating the sensitivity of the experiment to the $\beta$-delayed neutron decay of $^{132}$In.
The strong 450-keV transition feeds the 1226-keV level.
}
\end{figure}

\begin{figure*}
\centering
\includegraphics[width=0.677\textwidth]{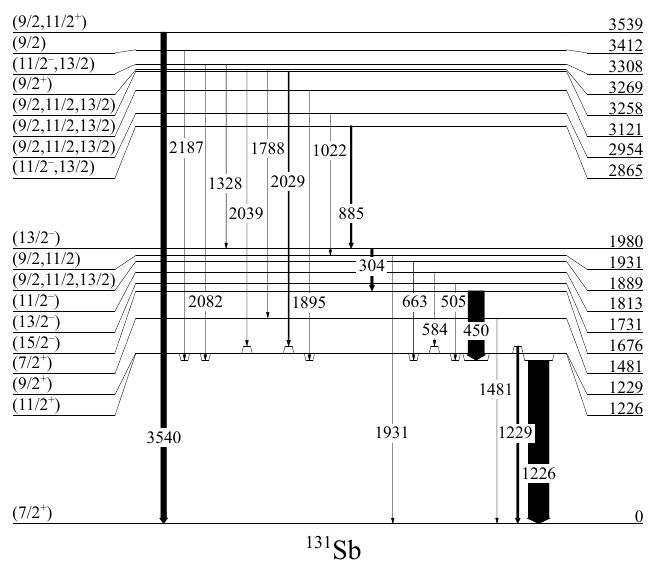}
\caption{\label{fig:131Sbls}
A level scheme showing the transitions observed in $^{131}$Sb following the $\beta$ decay of the high-spin $11/2^{-}$ isomer in $^{131}$Sn.
The widths of the arrows are proportional to the observed $\gamma$-ray intensity.
Energies are given in keV.
Level energies are taken from the ENSDF~\cite{Khazov2006}.
}
\end{figure*}

\begin{table*}
\caption{\label{tab:131Sb}
A summary of the excited states and $\gamma$-ray transitions observed in $^{131}$Sb following the $\beta$ decay of the high-spin $11/2^{-}$ isomer of $^{131}$Sn.
For each transition, the initial and final levels, $E_{x}$ and $E_{f}$, are listed along with the spins and parities $J^{\pi}_{i}\rightarrow J^{\pi}_{f}$ of the levels.
Finally, the relative branching ratios BR$_{\gamma}$ are compared to the ENSDF~\cite{Khazov2006}, and relative intensities I$_{\gamma}$ are compared to Ref.~\cite{Stone1988}.
Level energies are taken from the ENSDF~\cite{Khazov2006}.
}
\begin{tabular*}{\textwidth}{@{\extracolsep{\fill}}S S c S S S S S}
\hline \hline
{E$_{x}$\,(keV)} & {E$_{\gamma}$\,(keV)} & $J^{\pi}_{i}\rightarrow J^{\pi}_{f}$ & {E$_{f}$\,(keV)} & \multicolumn{2}{c}{BR$_{\gamma}$\,(rel.)} & \multicolumn{2}{c}{I$_{\gamma}$\,(rel.)} \\
\cline{5-6} \cline{7-8}
 & & & & {Present work} & {ENSDF} & {Present work} & {Ref.~\cite{Stone1988}} \\
\hline
1226.04(3)  & 1226.5(3) & $(11/2^{+})\rightarrow(7/2^{+})$       & 0       & 100    & 100     & 100    & 100 \\
1229.28(5)  & 1229.3(3) & $(9/2^{+})\rightarrow(7/2^{+})$        & 0       & 100    & 100(12) & 11(5)  & 20.2 \\
1481.15(3)  & 1481.5(5) & $(7/2^{+})\rightarrow(7/2^{+})$        & 0       & 100    & 100(4)  & 1.6(5)\footnote{Calculated from intensity of the 1788-keV transition} & \\
1676.06(6)  & 450.2(3)  & $(15/2^{-})\rightarrow(11/2^{+})$      & 1226.04 & 100    & 100(11) & 77(4)  & 72.6 \\
1730.6(3)   & 504.6(3)  & $(13/2^{-})\rightarrow(11/2^{+})$      & 1226.04 & 100    & 78(25)  & 3.3(6) & 2.66 \\
1812.93(13) & 583.8(3)  & $(11/2^{-})\rightarrow(9/2^{+})$       & 1229.28 & 100    & 100(19) & 2.3(5) & 4.9 \\
1889.5(7)   & 663(1)    & $(9/2,11/2,13/2)\rightarrow(11/2^{+})$ & 1226.04 & 100    &         & 1.1(4) & 2.6 \\
1931.08(8)  & 1931(1)   & $(9/2,11/2)\rightarrow(7/2^{+})$       & 0       & 100    & 100(22) & 3(1)   & 8.1 \\
1980.39(7)  & 304.4(5)  & $(13/2^{-})\rightarrow(15/2^{-})$      & 1676.06 & 100    & 100(13) & 12(2)\footnote{Calculated from intensities of the 885-keV and 1328-keV transitions} & 17.4 \\
2865.47(10) & 885.5(5)  & $(11/2^{-},13/2)\rightarrow(13/2^{-})$ & 1980.39 & 100    & 100(13)  & 8.6(7) & \\
2953.56(17) & 1021.7(4) & $(9/2,11/2,13/2)\rightarrow(9/2,11/2)$ & 1931.08 & 100    & 100(25)  & 1.8(3) & \\
3121.33(19) & 1894.6(4) & $(9/2,11/2,13/2)\rightarrow(11/2^{+})$ & 1226.04 & 100    & 88(20)   & 2.8(4) & \\
3258.58(18) & 2029.2(3) & $(9/2,11/2,13/2)\rightarrow(9/2^{+})$  & 1229.28 & 100    & 100(19)  & 4.8(6) & \\
3268.61(14) & 1787.9(4) & $(9/2^{+})\rightarrow(7/2^{+})$        & 1481.15 & 52(15) & 100(19)  & 1.6(5) & \\
            & 2039.0(4) & $(9/2^{+})\rightarrow(9/2^{+})$        & 1229.28 & 100    & 95(23)   & 3.1(5) & \\
3308.45(14) & 1328.1(4) & $(11/2^{-},13/2)\rightarrow(13/2^{-})$ & 1980.39 & 84(12) & 70(13)   & 3.1(4) & \\
            & 2082.4(5) & $(11/2^{-},13/2)\rightarrow(11/2^{+})$ & 1226.04 & 100    & 100(26)  & 3.7(5) & \\
3411.91(20) & 2186.9(5) & $(9/2)\rightarrow(11/2^{+})$           & 1226.04 & 100    & 118(30)  & 3.1(4) & \\
3539.1(10)  & 3540.0(3) & $(9/2,11/2^{+})\rightarrow(7/2^{+})$   & 0       & 100    & 100      & 26(3)  & \\
\hline \hline
\end{tabular*}
\end{table*}

\section{\label{sec:results} Results}

\subsection{\label{sec:pn} $\beta$-delayed neutron spectroscopy}

The $Q_{\beta}$-value from the ground state of $^{132}$In (14.140(60)~MeV) is higher than the neutron separation energy in $^{132}$Sn (7.353(4)~MeV)~\cite{Wang2017}, and $^{132}$In has been observed to exhibit $\beta$-delayed neutron emission into the nucleus $^{131}$Sn.
The probability of one-neutron emission, $P_{n}$, has been measured previously by counting neutrons and $\beta$ particles~\cite{Lund1980,Reeder1986,Rudstam1993} with varied results.
The current evaluated emission probability is 7.4(14)\%~\cite{Singh132In}, which is a weighted average of 6.8(14)\%~\cite{Reeder1986} and 10.7(33)\%~\cite{Rudstam1993}.
In the present experiment, it was possible to measure the $P_{n}$ by counting $\gamma$-rays in both $^{132}$Sn and the $A=131$ daughters.

Allowed $\beta$ decays from the $(7^{-})$ ground state in $^{132}$In will populate states in $^{132}$Sn with spin 6--8.
Subsequent neutron emission from a highly excited state will populate states in $^{131}$Sn with spin of at least 11/2.
If either the $\beta$ particle and antineutrino or the neutron carries away one unit of angular momentum, from a unique first-forbidden $\beta$ decay or $l=1$ neutron emission, then spins down to 9/2 are accessible.
In $^{131}$Sn, there is a low-spin $(3/2^{+})$ ground state and high-spin $(11/2^{-})$ isomer~\cite{Khazov2006}.
It is possible to populate this isomer with an allowed Gamow-Teller decay followed by the emission of an $l=0$ neutron.
No other known levels in $^{131}$Sn up to the $(15/2^{-})$ level located 4102~keV above the isomer have high enough spin to be significantly populated in the $\beta n$ decay, greatly reducing the energy available for neutron emission.

There are presently 19 $\gamma$-ray transitions listed in the ENSDF which feed either the $(3/2^{+})$ ground state or $(11/2^{-})$ isomer in $^{131}$Sn~\cite{Khazov2006}.
Of these, two transitions at 4220~keV and 4261~keV were identified in the current analysis.
The 4220-keV transition has been previously seen following fission of $^{248}$Cm and has been assigned to the decay of a $(13/2^{+})$ level to the $(11/2^{-})$ isomer~\cite{Bhattacharyya2001}.
The transition at 4261~keV has been identified from the $\beta$ decay of $^{131}$In~\cite{Dunlop2019,Fogelberg2004}.
It has been suggested to decay to the $(3/2^{+})$ ground state~\cite{Fogelberg2004}; however, the 4261-keV level has not been assigned any spin, and there are no coincident transitions.
Based on the current observation, it is more likely that this level has a high spin and decays to the $(11/2^{-})$ isomer.
The combined intensity of these two transitions yielded a total of $6.2(2)\times10^{4}$ $^{131}$Sn nuclei produced.
The number of $^{132}$Sn nuclei was calculated from the intensities of the three ground-state transitions (4041, 4352, and 4416~keV), giving $1.22(1)\times10^{7}$ total nuclei.
Comparing these numbers gives a $P_{n}$ of 0.50(2)\%.
Although the statisical error is relatively small, there is a larger systematic uncertainty due to several effects.
Most importantly, some of the transitions from high-spin levels, such as 4353 and 4423~keV, overlap with stronger transitions in $^{132}$Sn, and there are not enough counts to identify $\gamma$-$\gamma$ coincidences.
Additionally, there could be other unknown transitions which are not seen, and would be difficult to assign to $^{131}$Sn without a full coincidence analysis.
The measured 0.5\% branch should be taken as a lower limit, because any decay directly to the $(11/2^{-})$ isomer would not be detected with the current setup.
A $\beta n$ decay to the isomer via an intermediate $6^{-}$ level was tentatively proposed in Ref.~\cite{Bjornstad1986}, although the proposed intermediate level is now known to lie below the neutron separation energy.

\begin{figure}[t]
\includegraphics[width=\columnwidth]{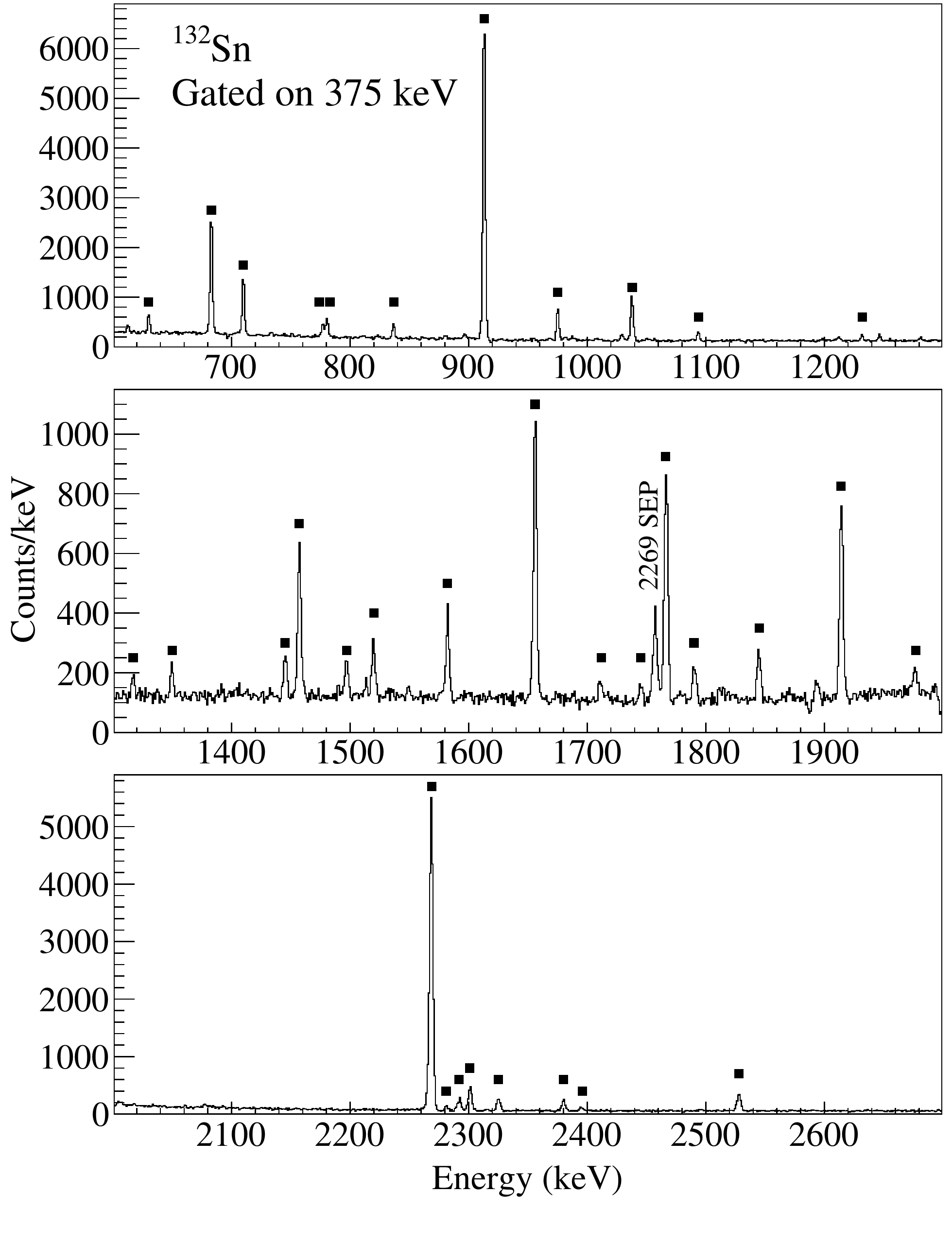}
\caption{\label{fig2}
A sample of the $\beta$-$\gamma$ spectrum gated on the 375-keV $4^{+}\rightarrow2^{+}$ transition in $^{132}$Sn.
The black squares highlight some of the coincident peaks which could be identified and placed in the level scheme.
}
\end{figure}

\begin{figure*}
\centering
\subfigure{\includegraphics[width=\textwidth]{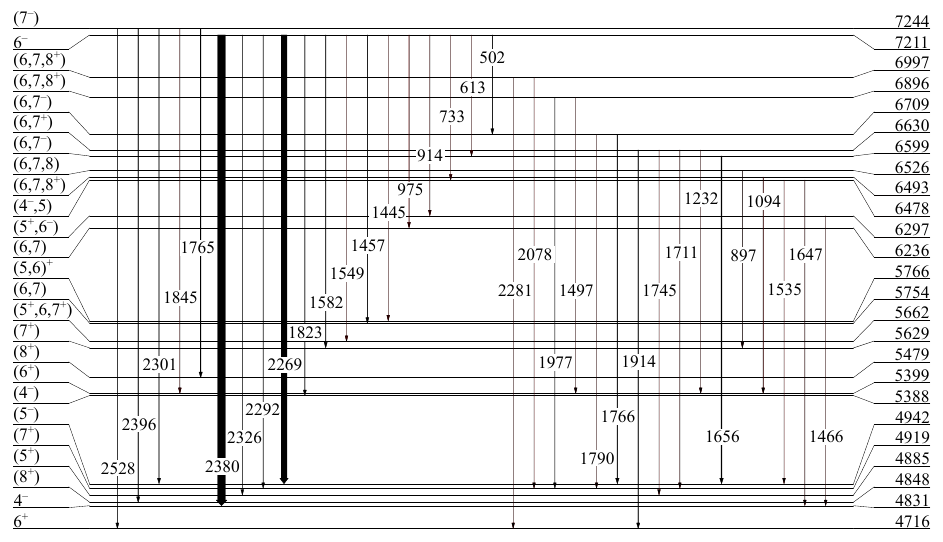}}
\vspace{0.1in}
\subfigure{\includegraphics[width=\textwidth]{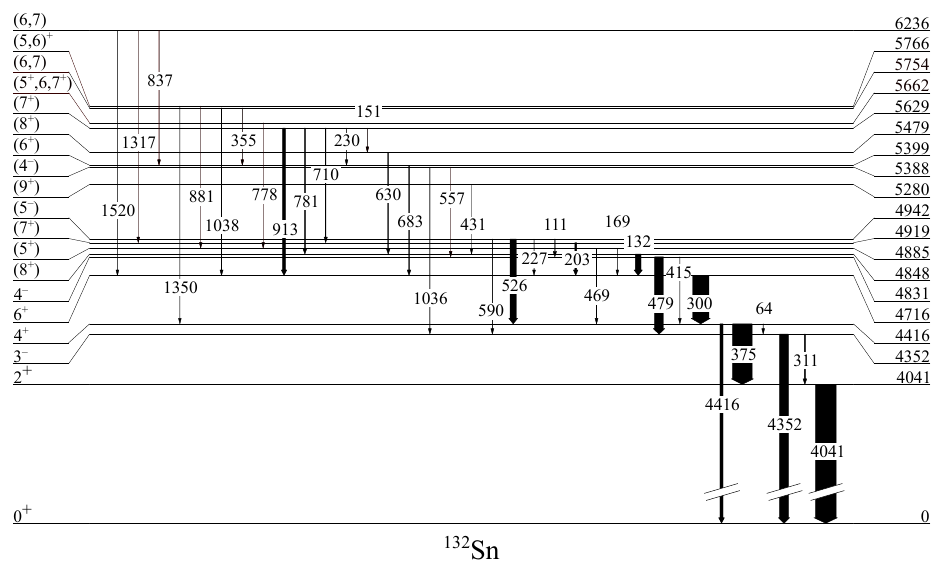}}
\caption{\label{fig:level}
A level scheme showing the transitions observed in $^{132}$Sn following the $\beta^{-}$ of $^{132}$In.
The widths of the arrows are proportional to the observed $\gamma$-ray intensity.
Energies are given in keV.
The two parts of the level scheme are not drawn to scale to each other.
}
\end{figure*}

Although the tape cycling was used to remove contamination from the beam, including long-lived decay products, the 37-s cycle lasted long enough for much of the $^{131}$Sn daughter nuclei ($T_{1/2}$=56.0(5)~s, 58.4(5)~s~\cite{Khazov2006}) to also $\beta$ decay at the experimental station.
Therefore it was possible to perform limited spectroscopy on $^{131}$Sb.
A sample $\gamma$-ray spectrum, gated on the 1226-keV $(11/2^{+})\rightarrow(7/2^{+})$ transition in $^{131}$Sb, is shown in Figure~\ref{fig:131Sbspec}.
The peak at 450~keV clearly indicates the presence of $^{131}$Sb.
A total of 19 $\gamma$-ray transitions belonging to $^{131}$Sb~\cite{Khazov2006} were observed in the experiment.
All of the transitions seen are displayed in the level scheme of Figure~\ref{fig:131Sbls}.
Relative $\gamma$-ray intensities and branching ratios are listed in Table~\ref{tab:131Sb}.
Almost all of the spin assignments for the currently observed levels are in the range $9/2\text{--}15/2$, indicating that these levels are much more likely to be fed from the $(11/2^{-})$ isomer of $^{131}$Sn, consistent with the expected $\beta n$ feeding patterns from $^{132}$In.

The number of $^{131}$Sb nuclei produced could be obtained from the observed intensities of the transitions.
In particular, the intensities of four out of the five transitions which feed the ground state could be measured directly in the $\gamma$ singles spectrum.
Only the intensity of the 1481-keV transition could not be measured in singles.
Direct $\beta$ feeding to the $(7/2^{+})$ 1481-keV level would constitute a unique first-forbidden transition, which is unlikely given the availability of higher-spin states.
Therefore it was assumed that this level had no direct $\beta$ feeding, and the intensity of the 1481-keV transition was matched to the 1788-keV transition above it.
Because the half-life of the isomer of $^{131}$Sn (58.4(5)~s~\cite{Khazov2006}) is longer than the decay portion of the tape cycle (5~s), it was necessary to make a correction to the observed intensities of those transitions to account for the fraction of nuclei which did not decay at the experimental station.
The total number of $^{131}$Sb nuclei produced was determined to be $1.72(4)\times10^{6}$, leading to a measured $P_{n}$ of 12.3(4)\%.
This value is much larger than the one obtained from the $\gamma$-ray transitions in $^{131}$Sn, and indicates that the $\beta n$ decay of $^{132}$In proceeds primarily through the $(11/2^{-})$ isomer in $^{131}$Sn.
This value is also larger than the current evaluated probability of 7.4(14)\%, although it is in reasonable agreement with the most recent $\beta-n$ counting experiment, which reported a value of 10.7(33)\%~\cite{Rudstam1993}.
In Ref.~\cite{Rudstam1993}, the $P_{n}$ values of several nuclei were tabulated.
Among the tabulated results, many were consistent with previous results, but in the case of $^{132}$In as well as a few others, the results were significantly larger. 
There was no general trend to measure larger values, and it is not clear why the previous measurements were smaller for the case of $^{132}$In.

\subsection{\label{sec:131spectroscopy} Spectroscopy of $^{131}$Sb}

New spin and parity assignments have been determined for several known levels in $^{131}$Sb.
In particular, the 1889, 2865, 2954, 3121, 3258, 3269, and 3539-keV levels were assumed to be directly fed from the $\beta$ decay of the $(11/2^{-})$ isomer of $^{131}$Sn, limiting the spins to (9/2,11/2,13/2) for allowed or first-forbidden transitions.
Several of these spins and parities can be restricted further based on $\gamma$-ray feeding patterns.
The 2865-keV level was previously assigned a tentative spin of (11/2,13/2,15/2) based on $\gamma$-ray feeding patterns, including a weak transition to 1676-keV $(15/2^{-})$ level.
Thus this level is assigned a tentative spin of $(11/2^{-},13/2)$.
The 3269-keV level was previously assigned a tentative spin of $(5/2\text{--}9/2)$ based on $\gamma$-ray feeding patterns.
The current observation of direct $\beta$ feeding of this level suggests a spin of $(9/2^{+})$.
The level at 3308~keV was observed to decay to the 1226-keV $(11/2^{+})$ and 1980-keV $(13/2^{-})$ levels.
Previously a weak transition to the 1916-keV $(15/2^{-})$ level was also observed~\cite{Khazov2006}.
As a result, the 3308-keV level is assigned a tentative spin of $(11/2^{-},13/2)$.
The 3539-keV state is observed to decay to the $(7/2^{+})$ ground state, restricting this state to $(9/2,11/2^{+})$.
Finally, the state at 1931~keV was previously given a tentative spin assignment of $(7/2\text{--}11/2)$ based on $\gamma$-ray feeding patterns.
The intensities of the 1931-keV and 1022-keV transitions indicate that there is a small amount of direct $\beta$ feeding to this state, restricting the spin to $(9/2,11/2)$.

\subsection{\label{sec:132spectroscopy} Spectroscopy of $^{132}$Sn}

A total of $5\times10^{8}$ $\gamma$-addback events and $1.1\times10^{8}$ $\gamma$-$\gamma$ coincidence events were observed in the analysis.
Based on the $\gamma$-$\gamma$ statistics, a total of 70 transitions have been placed along with 29 excited states.
A sample of the $\beta$-coincident spectrum gated on the 375-keV $4^{+}\rightarrow2^{+}$ transition is shown in Figure~\ref{fig2}.
A total of 53 transitions were observed in coincidence with this gate, with relative intensities down to 0.14\%.
The $\gamma$-$\gamma$ coincidence analysis was used to construct the level scheme which can be seen in Figure~\ref{fig:level}.

A list of all observed excited states and $\gamma$-ray transitions with their measured intensities is shown in Table~\ref{table1}.
Intensities could be determined directly from the addback singles spectrum for 41 of the 70 transitions, while the intensities for the remaining 29 transitions were determined from gated spectra.
The $\gamma$ intensities are reported relative to the 4.041-MeV $2^{+}\rightarrow0^{+}$ transition.
Because the efficiency at 4.041~MeV was tied to the efficiency in the 300-keV region, the relative uncertainty on the efficiency at this energy could be limited to 0.4\%, even though the relative uncertainty on the extrapolated source efficiency curve was much larger above 4~MeV.
Therefore, for strong peaks with a statistical uncertainty of $\lesssim0.4\%$, the uncertainty on the relative intensity is less than 1\%.
In Table~\ref{table1}, $\gamma$ intensities and branching ratios are compared to the evaluated values.
It is difficult to determine the level of agreement for many of the transitions, because the uncertainties in the intensities were not reported by Fogelberg \emph{et al.}~\cite{Fogelberg1994}.
Instead, errors listed in the table were assigned by the evaluators.
However, there is generally good agreement between the current and evaluated intensities.

Prior to the work of Ref.~\cite{Benito2020}, the most comprehensive $\beta$-decay study performed placed 21 excited states and 44 transitions in the level scheme~\cite{Fogelberg1994}.
Of these, 20 excited states and 42 transitions have been confirmed in the current work.
The 70-keV transition which feeds the $8^{+}$ level at 4848~keV was not seen.
Another transition at 774~keV and the level it depopulates at 6173~keV could not be confirmed.
However, both the 1457-keV and 1038-keV transitions previously associated with this level were observed.
The observation of a 355-keV transition in coincidence with the 1457~keV transition suggests a reordering of the 1457 and 1038-keV transitions and the placement of a level at 5754~keV.
A possible 89-keV transition suggested by Bj\"{o}rnstad \emph{et al.}~\cite{Bjornstad1986} to decay from the level at 4919~keV was seen neither by Fogelberg \emph{et al.}~\cite{Fogelberg1994} nor in the current work.

A 431-keV transition from a level at 5280~keV was reported to be in delayed coincidence with the 132-keV $(8^{+})\rightarrow(6^{+})$ transition from a fission experiment~\cite{Bhattacharyya2001}.
This transition was assigned to be from a ($9^{+}$) based on systematic comparisons with $^{208}\mathrm{Pb}$.
This level was observed in the current experiment, and the $\beta$ feeding to this level was measured.

A recent experiment on the $\beta n$ decay of $^{133}\mathrm{In}$~\cite{Piersa2019} also observed several new transitions at 1030, 1350, and 1374~keV.
The 1350-keV transition and the 5766-keV level which it depopulates were confirmed in the present work.
The placement of this level was supported by the additional observation of a 1445-keV transition which feeds it from the 7211-keV $6^{-}$ level, as well as an 881-keV transition which decays from this level to the 4885-keV $(5^{+})$ level.
Neither the 1030-keV transition nor the 1374-keV transition could be confirmed.

From the measured $\gamma$-ray intensities, the apparent direct $\beta$ feeding to excited states in $^{132}$Sn could be calculated.
The feeding calculations take into account theoretical conversion coefficients, calculated using the BrIcc code~\cite{Kibedi2008} with the frozen orbital approximation.
Direct feeding could be observed into 22 of the 30 levels, with upper limit placed on additional four levels.
Based on the feeding intensities, $\log ft$ values could be calculated for the excited states.
The feeding intensities and $\log ft$ values are listed in Table~\ref{table2}.
Intensities have been calculated with a 12.3\% $\beta$-delayed neutron decay branch.

Spin and parity assignments for levels were primarily based on $\gamma$ feeding patterns, as well as $\log ft$ values.
The spins of several excited state with previous assignments have been restricted based on the observation of additional $\gamma$-rays.
The level at 5766~keV was not previously assigned any spin.
In the present work, the spins and parity of this level was determined to be $(5,6)^{+}$ based on the observation of $\gamma$-ray transitions to the 4885-keV $(5^{+})$ and 4416-keV $4^{+}$ levels and a $\log ft$ value which is consistent with first-forbidden or unique first-forbidden decay.
The level at 6236~keV was previously given a range of spins of $(6,7,8^{+})$, based on $\gamma$-ray feeding patterns.
The presently observed 975-keV transition feeds this level from the $(6^{-})$ level at 7211~keV, which eliminates an $8^{+}$ assignment, so the reported spin is (6,7).
The level at 6630~keV was also given a range of spins of $(6,7,8^{+})$ based on $\gamma$-ray feeding patterns.
Several transitions from this state have been observed, including a 1745-keV transition which connects this level to the $(5^{+})$ level at 4885~keV.
This rules out the possibility of an $8^{+}$ or $7^{-}$ spin, and the current reported spin is $(6,7^{+})$.
The level at 6896~keV was assigned a range of (6,7,8) based on $\gamma$-ray feeding patterns.
The presently observed 1497-keV transition to the $(6^{+})$ level at 5398~keV restricts this level to $(6,7,8^{+})$.

Spins and parities have also been tentatively assigned for other levels identified in the current work.
The level at 5662~keV is fed from the 7211-keV $6^{-}$ level and decays to the 4885-keV $(5^{+})$ level.
It also has a small amount of direct $\beta$ feeding, and the $\log ft$ value is consistent with first-forbidden or unique first-forbidden decay.
Therefore, this level is assigned a tentative spin of $(5^{+},6,7^{+})$.
The level at 5754~keV is fed from the 7211-keV $6^{-}$ level and decays to two $6^{+}$ levels; the $\log ft$ value is consistent with an allowed or first-forbidden transition.
This level is therefore assigned a tentative spin of (6,7).
The 6297-keV level is fed from the 7211-keV $6^{-}$ level and decays to the 4831-keV $(4^{-})$ level.
The $\log ft$ value is consistent with either an allowed or unique first-forbidden transition, and this level is assigned a tentative spin of $(5^{+},6^{-})$.
The level at 6478~keV is fed from the 7211-keV $6^{-}$ level and decays to the 4831-keV $4^{-}$ and 4942-keV $(5^{-})$ levels.
It has a $\beta$ feeding component with a two-sigma limit $<0.04\%$, suggesting that it is not an allowed or first-forbidden transition.
This level is assigned a tentative spin of $(4^{-},5)$.
Three levels at 6493, 6526, and 6997~keV are directly fed from the $\beta$ decay of $^{132}$In.
The $\log ft$ values for these levels are consistent with allowed or first-forbidden transitions, limiting the spins to (6,7,8).
In addition, the 6493 and 6997-keV levels decay to known $(6^{+})$ levels, further restricting their spins to $(6,7,8^{+})$.

\begin{ThreePartTable}
\renewcommand\TPTminimum{\textwidth}
\begin{TableNotes}
\item[a] \label{tna} The evaluator assigned 5\% uncertainty for $\mathrm{I}_{\gamma}>10$ values, 10\% for $\mathrm{I}_{\gamma}=3\text{--}10$ and 15\% for $\mathrm{I}_{\gamma}<3$.
\item[b] \label{tnb} Relative intensities have been multiplied by 100/107.
\item[c] \label{tnc} Spins and parities determined in present work. See text for details.
\item[d] \label{tnd} Transition re-placed in the level scheme. See text for details.
\end{TableNotes}

\begin{longtable*}{@{\extracolsep{\fill}}S S c S S S S S}
\caption{\label{table1}
A summary of the excited states and $\gamma$-ray transitions observed in $^{132}$Sn following the $\beta$ decay of $^{132}$In.
For each transition, the initial and final levels, $E_{x}$ and $E_{f}$, are listed along with the spins and parities $J^{\pi}_{i}\rightarrow J^{\pi}_{f}$ of the levels.
Finally, the relative branching ratios BR$_{\gamma}$ and relative intensities I$_{\gamma}$ are compared to the ENSDF~\cite{Singh132Sn}.
} \\
\hline \hline
{$E_{x}$\,(keV)} & {$E_{\gamma}$\,(keV)} & $J^{\pi}_{i}\rightarrow J^{\pi}_{f}$ & {$E_{f}$\,(keV)} & \multicolumn{2}{c}{BR$_{\gamma}$\,(rel.)} & \multicolumn{2}{c}{I$_{\gamma}$\,(rel.)} \\
\cline{5-6} \cline{7-8}
 & & & & {Present work} & {ENSDF}\tnotex{tna} & {Present work} & {ENSDF\tnotex{tna} \tnotex{tnb}} \\
\hline
\endfirsthead
\caption[]{\em{(Continued).}} \\
\hline \hline
{E$_{x}$\,(keV)} & {E$_{\gamma}$\,(keV)} & $J^{\pi}_{i}\rightarrow J^{\pi}_{f}$ & {E$_{f}$\,(keV)} & \multicolumn{2}{c}{BR$_{\gamma}$\,(rel.)} & \multicolumn{2}{c}{I$_{\gamma}$\,(rel.)} \\
\cline{5-6} \cline{7-8}
 & & & & {Present work} & {ENSDF} & {Present work} & {ENSDF} \\
\hline
\endhead
\hline
\endfoot
\hline \hline
\insertTableNotes
\endlastfoot
4041.2(2)             & 4041.24(20) & $2^{+}\rightarrow0^{+}$             & 0      & 100     & 100    & 100      & 100(5) \\
4351.9(2)             & 310.74(20)  & $3^{-}\rightarrow2^{+}$             & 4041.1 & 11.7(3) & 11(1)  & 4.9(1)   & 4.7(5) \\
                      & 4351.88(21) & $3^{-}\rightarrow0^{+}$             & 0      & 100     & 100    & 41.8(4)  & 42(2) \\
4416.3(2)             & 64.4(2)     & $4^{+}\rightarrow3^{-}$             & 4351.9 & 0.8(4)  & 1.3(2) & 0.8(4)   & 1.2(2) \\
                      & 375.07(20)  & $4^{+}\rightarrow2^{+}$             & 4041.1 & 100     & 100    & 93.4(8)  & 93(5) \\
                      & 4416.22(20) & $4^{+}\rightarrow0^{+}$             & 0      & 15.8(2) & 17(1)  & 14.7(2)  & 15.6(8) \\
4715.9(3)             & 299.59(20)  & $6^{+}\rightarrow4^{+}$             & 4416.3 & 100     & 100    & 78.7(6)  & 73(4) \\
4830.9(3)             & 414.66(21)  & $4^{-}\rightarrow4^{+}$             & 4416.3 & 1.76(6) & 2.1(5) & 0.76(3)  & 0.8(2) \\
                      & 479.03(20)  & $4^{-}\rightarrow3^{-}$             & 4351.9 & 100     & 100    & 42.9(4)  & 40(2) \\
4848.4(4)             & 132.49(22)  & $(8^{+})\rightarrow6^{+}$           & 4715.9 & 100     & 100    & 27.3(3)  & 22(1) \\
4885.4(3)             & 169.4(3)    & $(5^{+})\rightarrow6^{+}$           & 4715.9 & 10(3)   & 20(7)  & 0.16(5)  & 0.28(9) \\
                      & 469.15(21)  & $(5^{+})\rightarrow4^{+}$           & 4416.3 & 100     & 100    & 1.63(4)  & 1.4(2) \\
4919.0(3)             & 203.12(20)  & $(7^{+})\rightarrow6^{+}$           & 4715.9 & 100     & 100    & 8.4(3)   & 7.0(7) \\
4942.4(3)             & 111.43(20)  & $(5^{-})\rightarrow4^{-}$           & 4830.9 & 11.0(2) & 9(1)   & 3.19(5)  & 2.7(3) \\
                      & 226.64(21)  & $(5^{-})\rightarrow6^{+}$           & 4715.9 & 2.7(2)  & 2.8(6) & 0.78(4)  & 0.8(2) \\
                      & 526.12(20)  & $(5^{-})\rightarrow4^{+}$           & 4416.3 & 100     & 100    & 29.1(3)  & 30(2) \\
                      & 590.45(21)  & $(5^{-})\rightarrow3^{-}$           & 4351.9 & 4.5(1)  & 7(1)   & 1.31(4)  & 2.0(3) \\
5279.7(5)             & 431.3(3)    & $(9^{+})\rightarrow(8^{+})$         & 4848.4 & 100     &        & 0.4(1)   & \\
5387.8(3)             & 556.8(3)    & $(4^{-})\rightarrow4^{-}$           & 4830.9 & 12.4(6) &        & 0.15(2)  & \\
                      & 1035.71(22) & $(4^{-})\rightarrow3^{-}$           & 4351.9 & 100     & 100    & 1.18(6)  & 1.3(2) \\
5398.9(4)             & 682.96(21)  & $(6^{+})\rightarrow6^{+}$           & 4715.9 & 100     & 100    & 4.60(8)  & 4.9(5) \\
5478.7(4)             & 630.24(20)  & $(8^{+})\rightarrow(8^{+})$         & 4848.4 & 100     & 100    & 5.02(6)  & 6.4(7) \\
5628.9(3)             & 150.6(4)    & $(7^{+})\rightarrow(8^{+})$         & 5478.7 & 1.9(1)  &        & 0.25(2)  & \\
                      & 230.03(22)  & $(7^{+})\rightarrow(6^{+})$         & 5398.9 & 6.8(3)  & 7(2)   & 0.92(4)  & 0.9(2) \\
                      & 709.95(21)  & $(7^{+})\rightarrow(7^{+})$         & 4919.0 & 18.1(4) & 23(3)  & 2.45(4)  & 3.0(4) \\
                      & 780.59(21)  & $(7^{+})\rightarrow(8^{+})$         & 4848.4 & 37.2(6) & 29(4)  & 5.04(6)  & 3.8(4) \\
                      & 913.08(20)  & $(7^{+})\rightarrow6^{+}$           & 4715.9 & 100     & 100    & 13.6(1)  & 13.1(9) \\
5662.4(4)\tnotex{tnc} & 777.96(21)  & $(5^{+},6,7^{+})\rightarrow(5^{+})$ & 4885.4 & 100     &        & 0.17(1)  & \\
5754.2(4)\tnotex{tnc} & 355.01(25)  & $(6,7)\rightarrow(6^{+})$           & 5398.9 & 18(2)   &        & 0.34(3)  & \\
                      & 1037.76(21)\tnotex{tnd} & $(6,7)\rightarrow6^{+}$ & 4715.9 & 100     &        & 1.84(4)  & 1.4(2) \\
5766.0(3)\tnotex{tnc} & 881.1(3)    & $(5,6)^{+}\rightarrow(5^{+})$       & 4885.4 & 38(5)   &        & 0.095(8) & \\ 
                      & 1349.75(22) & $(5,6)^{+}\rightarrow4^{+}$         & 4416.3 & 100     &        & 0.25(3)  & \\ 
6236.0(4)\tnotex{tnc} & 836.8(3)    & $(6,7)\rightarrow(6^{+})$           & 5398.9 & 59(4)   &        & 0.48(2)  & \\
                      & 1317.0(3)   & $(6,7)\rightarrow(7^{+})$           & 4919.0 & 17(3)   &        & 0.14(2)  & \\
                      & 1520.0(3)   & $(6,7)\rightarrow6^{+}$             & 4715.9 & 100     & 100    & 0.81(4)  & 0.56(9) \\
6297.3(5)\tnotex{tnc} & 1466.4(4)   & $(5^{+},6^{-})\rightarrow4^{-}$     & 4830.9 & 100     &        & 0.16(3)  & \\
6477.7(5)\tnotex{tnc} & 1646.5(4)   & $(4^{-},5)\rightarrow4^{-}$         & 4830.9 & 100     &        & 0.25(3)  & \\
                      & 1535.1(5)   & $(4^{-},5)\rightarrow(5^{-})$       & 4942.4 & 22(7)   &        & 0.05(2)  & \\
6492.6(5)             & 1093.7(3)   & $(6,7,8^{+})\rightarrow(6^{+})$     & 5398.9 & 100     &        & 0.45(2)  & \\
6526.1(5)             & 897.2(3)    & $(6,7,8)\rightarrow(7^{+})$         & 5628.9 & 100     &        & 0.26(2)  & \\
6598.5(4)             & 1656.05(21) & $(6,7^{-})\rightarrow(5^{-})$       & 4942.4 & 100     & 100    & 3.88(6)  & 3.8(4) \\
6630.3(4)\tnotex{tnc} & 1231.8(3)   & $(6,7^{+})\rightarrow(6^{+})$       & 5398.9 & 10.9(7) &        & 0.24(1)  & \\
                      & 1711.4(3)   & $(6,7^{+})\rightarrow(7^{+})$       & 4919.0 & 10.6(6) &        & 0.24(1)  & \\
                      & 1745.1(3)   & $(6,7^{+})\rightarrow(5^{+})$       & 4885.4 & 6.2(6)  &        & 0.14(1)  & \\
                      & 1914.45(21) & $(6,7^{+})\rightarrow6^{+}$         & 4715.9 & 100     & 100    & 2.22(8)  & 2.1(3) \\
6709.1(4)             & 1766.59(21) & $(6,7^{-})\rightarrow(5^{-})$       & 4942.4 & 100     & 100    & 3.01(7)  & 4.1(5) \\
                      & 1789.9(3)   & $(6,7^{-})\rightarrow(7^{+})$       & 4919.0 & 8.8(5)  &        & 0.26(1)  & \\
6895.9(5)\tnotex{tnc} & 1497.2(3)   & $(6,7,8^{+})\rightarrow(6^{+})$     & 5398.9 & 100     &        & 0.38(4)  & \\
                      & 1976.9(3)   & $(6,7,8^{+})\rightarrow(7^{+})$     & 4919.0 & 94(18)  & 100    & 0.35(6)  & 0.37(9) \\
6997.2(4)\tnotex{tnc} & 2078.2(3)   & $(6,7,8^{+})\rightarrow(7^{+})$     & 4919.0 & 45(8)   &        & 0.21(3)  & \\
                      & 2281.4(3)   & $(6,7,8^{+})\rightarrow6^{+}$       & 4715.9 & 100     &        & 0.46(3)  & \\
7211.1(4)             & 501.81(22)  & $6^{-}\rightarrow(6,7^{-})$         & 6709.1 & 2.7(1)  & 2.9(5) & 1.01(4)  & 1.1(2) \\
                      & 612.7(3)    & $6^{-}\rightarrow(6,7^{-})$         & 6598.5 & 0.92(5) &        & 0.35(2)  & \\
                      & 733.4(3)    & $6^{-}\rightarrow(4^{-},5)$         & 6477.7 & 0.83(8) &        & 0.32(3)  & \\
                      & 913.8(3)    & $6^{-}\rightarrow(5^{+},6^{-})$     & 6297.3 & 0.17(1) &        & 0.063(6) & \\
                      & 975.0(3)    & $6^{-}\rightarrow(6,7)$             & 6236.0 & 2.11(7) &        & 0.80(2)  & \\
                      & 1445.34(25) & $6^{-}\rightarrow(5,6)^{+}$         & 5766.0 & 0.61(4) &        & 0.23(1)  & \\
                      & 1457.24(22)\tnotex{tnd} & $6^{-}\rightarrow(6,7)$ & 5754.2 & 3.7(1)  &        & 1.39(4)  & 1.4(2) \\
                      & 1548.6(3)   & $6^{-}\rightarrow(5^{+},6,7^{+})$   & 5662.4 & 0.28(1) &        & 0.106(6) & \\
                      & 1582.22(21) & $6^{-}\rightarrow(7^{+})$           & 5628.9 & 3.2(1)  & 3.1(2) & 1.21(4)  & 1.2(2) \\
                      & 1823.1(3)   & $6^{-}\rightarrow(4^{-})$           & 5387.8 & 3.9(2)  & 3.1(5) & 1.49(8)  & 1.2(2) \\
                      & 2268.62(21) & $6^{-}\rightarrow(5^{-})$           & 4942.4 & 71(1)   & 67(5)  & 27.2(4)  & 26(1) \\
                      & 2292.1(3)   & $6^{-}\rightarrow(7^{+})$           & 4919.0 & 2.62(9) & 3.1(5) & 1.00(3)  & 1.2(2) \\
                      & 2325.6(3)   & $6^{-}\rightarrow(5^{+})$           & 4885.4 & 2.35(8) & 1.9(5) & 0.90(3)  & 0.8(2) \\
                      & 2380.24(20) & $6^{-}\rightarrow4^{-}$             & 4830.9 & 100     & 100    & 38.1(5)  & 39(2) \\
7243.9(4)             & 1764.93(21) & $(7^{-})\rightarrow(8^{+})$         & 5478.7 & 100     & 88(17) & 2.51(4)  & 2.0(3) \\
                      & 1844.7(3)   & $(7^{-})\rightarrow(6^{+})$         & 5398.9 & 19.3(6) &        & 0.48(1)  & \\
                      & 2301.33(21) & $(7^{-})\rightarrow(5^{-})$         & 4942.4 & 92(5)   & 79(16) & 2.3(1)   & 1.8(3) \\
                      & 2395.6(3)   & $(7^{-})\rightarrow(8^{+})$         & 4848.4 & 88(2)   & 100    & 2.21(5)  & 2.2(3) \\
                      & 2528.21(21) & $(7^{-})\rightarrow6^{+}$           & 4715.9 & 59(2)   & 75(16) & 1.49(4)  & 1.7(3) \\
\end{longtable*}
\end{ThreePartTable}

\begin{table}
\caption{\label{table2}
Calculated $\beta$ feeding intensities and $\log ft$ values for the excited states in $^{132}$Sn from the $\beta$ decay of the $(7^{-})$ ground state of $^{132}$In.
Feeding calculations take into account theoretical conversion coefficients, calculated using the BrIcc code~\cite{Kibedi2008} with the frozen orbital approximation.
Values from the current work are calculated with a 12.3\% $\beta$-delayed neutron decay branch.
Limits for the current intensities are given at the two-sigma level.
ENSDF values were calculated by the evaluators from the level scheme of Fogelberg \emph{et al.}~\cite{Fogelberg1994} with a 7.4\% $\beta$-delayed neutron branch and assuming 5--15\% errors on $\gamma$ intensities.
Where different $\log ft$ values are possible based on spin assignments, unique transitions are listed below non-unique transitions.
}
\begin{tabular*}{\columnwidth}{@{\extracolsep{\fill}}c c S[table-format = <1.2] S S[table-format = <2.2] S}
\hline \hline
{$E_{x}$ (keV)} & {$J^{\pi}$} & \multicolumn{2}{c}{$I_{\beta}$ (\%)} & \multicolumn{2}{c}{$\log ft$} \\
\cline{3-4} \cline{5-6}
 & & {Present} & {ENSDF} & {Present} & {ENSDF} \\
\hline
4041 & $2^{+}$           & 0        & 0       &         & \\
4352 & $3^{-}$           & 0        & 0       &         & \\
4416 & $4^{+}$           & 0        & 0       &         & \\
4716 & $6^{+}$           & 1.3(4)   & 5(3)    & 6.7(1)  & 6.1(3) \\
4831 & $4^{-}$           & <1.1     & 0       & >11.5\footnote{\label{usf}Unique second-forbidden transition}   & \\
4848 & $(8^{+})$         & 17.4(2)  & 13(2)   & 5.57(2) & 5.7(1) \\
4885 & $(5^{+})$         & 0.29(4)  & 0.6(2)  & 9.65(7)\footnote{\label{uff}Unique first-forbidden transition} & 9.3(2) \\
4919 & $(7^{+})$         & 2.5(2)   & 2.0(5)  & 6.39(3) & 6.5(1) \\
4942 & $(5^{-})$         & <0.1     &         & >7.8    & \\
5280 & $(9^{+})$         & 0.23(6)  &         & 9.6(1)\footref{uff} & \\
5388 & $(4^{-})$         & <0.02    & 0       & >13\footref{usf} & \\
5399 & $(6^{+})$         & 0.70(6)  & 2.1(4)  & 6.84(4) & 6.4(1) \\
5479 & $(8^{+})$         & 1.24(4)  & 2.6(5)  & 6.57(2) & 6.3(1) \\
5629 & $(7^{+})$         & 11.7(1)  & 11.5(8) & 5.57(2) & 5.57(4) \\
5662 & $(5^{+},6,7^{+})$ & 0.033(7) &         & 8.1(1)  & \\
     &                   &          &         & 10.4(1)\footref{uff} & \\
5754 & $(6,7)$           & 0.45(4)  &         & 6.95(5) & \\
5766 & $(5,6)^{+}$       & 0.07(2)  &         & 7.8(2)  & \\
     &                   &          &         & 10.0(1)\footref{uff} & \\
6236 & $(6,7)$           & 0.35(8)  & 0.33(6) & 6.9(1)  & 7.0(1) \\
6297 & $(5^{+},6^{-})$   & 0.06(2)  &         & 7.7(1)  & \\
     &                   &          &         & 9.9(1)\footref{uff} & \\
6478 & $(4^{-},5)$       & <0.04    &         & >10\footref{uff} & \\
     &                   &          &         & >12\footref{usf} & \\
6493 & $(6,7,8^{+})$     & 0.25(1)  &         & 7.02(2) & \\
6526 & $(6,7,8)$         & 0.148(9) &         & 7.24(4) & \\
6599 & $(6,7^{-})$       & 1.98(4)  & 2.3(3)  & 6.10(2) & 6.0(1) \\
6630 & $(6,7^{+})$       & 1.59(5)  & 1.2(2)  & 6.19(2) & 6.3(1) \\
6709 & $(6,7^{-})$       & 1.27(5)  & 1.8(3)  & 6.26(2) & 6.1(1) \\
6896 & $(6,7,8^{+})$     & 0.41(4)  & 0.22(6) & 6.71(5) & 7.0(1) \\
6997 & $(6,7,8^{+})$     & 0.38(2)  &         & 6.71(3) & \\
7211 & $6^{-}$           & 41.6(4)  & 42(2)   & 4.61(2) & 4.61(3) \\
7244 & $(7^{-})$         & 5.05(8)  & 4.5(4)  & 5.52(2) & 5.57(5) \\
\hline \hline
\end{tabular*}
\end{table}

\subsection{\label{sec:halflife} Half-life of $^{132}$In}

\begin{figure}
\includegraphics[width=\columnwidth]{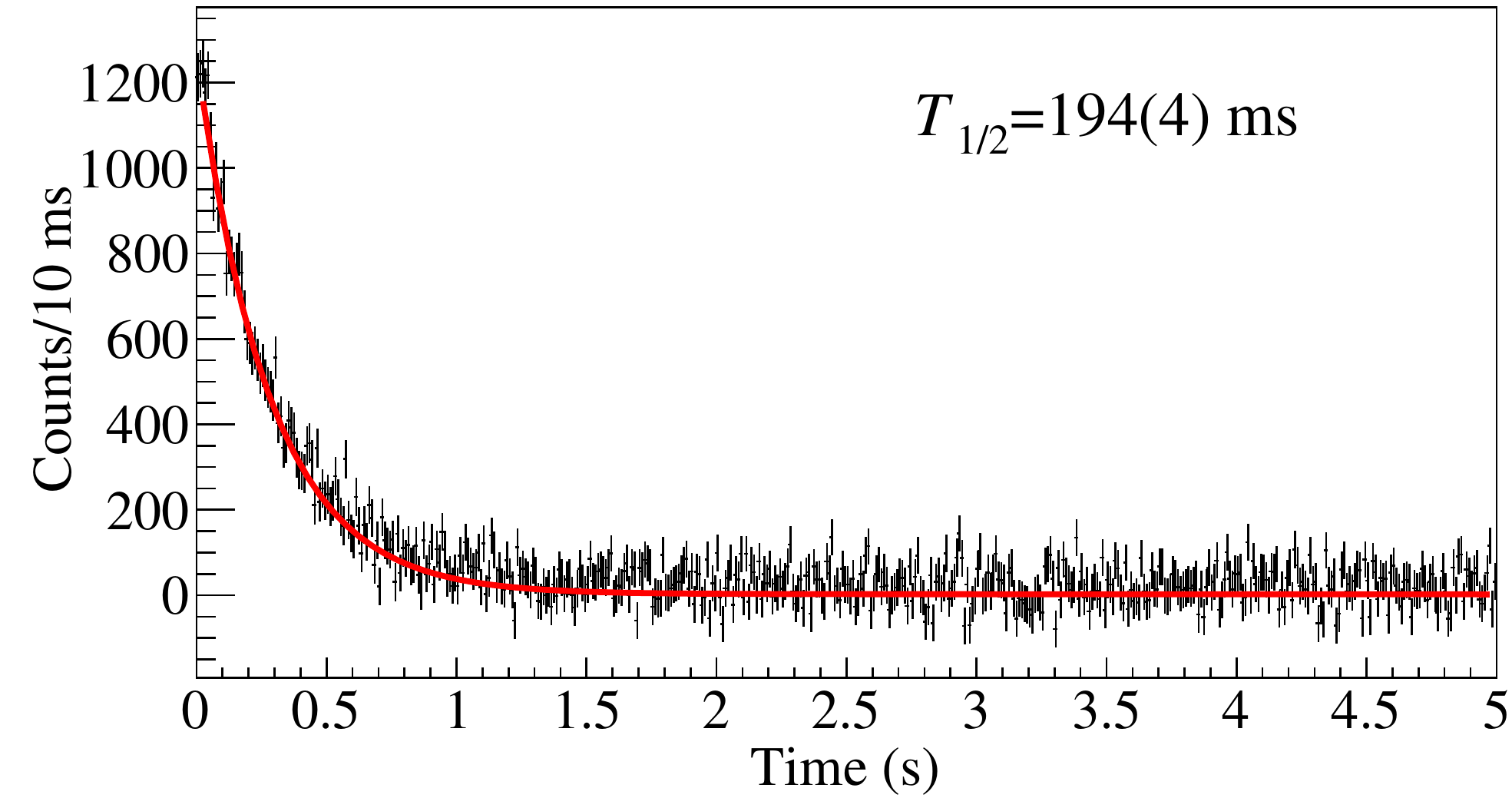}
\caption{\label{fig3}
The lifetime measurement for the decay of  the $(7^{-})$ ground state in $^{132}$In.
Shown is a background-subtracted sum of gates on the 300, 375, 2269, 2380, 4041, 4352, and 4416-keV transitions in $^{132}$Sn, plotted as a function of the time since the beam was blocked.
The best-fit curve is shown in red, which gives a half-life of 194(4)~ms.
}
\end{figure}

The half-life of $^{132}$In was measured by plotting the $\gamma$-ray counts as a function of time and placing gates on several of the strongest transitions in $^{132}$Sn.
Gates were placed on intense transitions at 300, 375, 2269, 2380, 4041, 4352, and 4416~keV, with background subtraction for each gate.
The results are shown in Figure~\ref{fig3}.
The decay was fit with a simple exponential function $f(t)=A+Be^{-\lambda t}$.
The best-fit half-life based on the $\chi^2$ of the fit is $T_{1/2}$~=~194(4)~ms.
A ``chop-analysis'' was performed by adjusting the range of the fit region to take into account possible rate-dependent effects.
The result is a small increase in the systematic error.
Therefore the reported value of the half-life is 194(5)~ms.
This agrees with the evaulated value of 200(2)~ms~\cite{Singh132In}.

\subsection{\label{sec:AC} Angular correlations}

It was possible to perform angular correlations using the $\gamma$-ray cascades in $^{132}$Sn.
Angular correlations have the general form
\begin{equation}
\label{eq1}
W(\theta)=A_{00}\left[1+a_{2}P_{2}(\cos\theta)+a_{4}P_{4}(\cos\theta)\right].
\end{equation}
Here, $\theta$ is the angle between successive $\gamma$ rays emitted in cascade, $A_{00}$ is a normalization factor, $a_{2}$ and $a_{4}$ are coefficients which depend on the multipolarities and mixing ratios of the $\gamma$-ray transitions, and $P_{i}(\cos\theta)$ are the Legendre polynomials.

Details about angular correlation analysis with GRIFFIN can be found in Ref.~\cite{Smith2019}.
A similar analysis obtaining angular correlations with the current experimental setup has been performed in Ref.~\cite{Ortner2020} and with a different setup in Ref.~\cite{Pore2019}.
For the angular correlations, clover addback was not used; instead, photopeaks were measured using the 64 single crystals.
The geometry of the HPGe detectors within GRIFFIN creates 51 distinct angles between pairs of detector crystals, ranging from 19\textdegree\ to 180\textdegree.
In the present analysis, the 51 angles were grouped into 22 points in order to increase statistics for each data point.
Angles used in the grouped analysis were calculated as a weighted average of the original angular pairs; each grouped angle had between one and five original angles.
The data were also folded around 90\textdegree\ because of the symmetry of the correlations, resulting in 11 points.

\begin{figure}[t]
\includegraphics[width=\columnwidth]{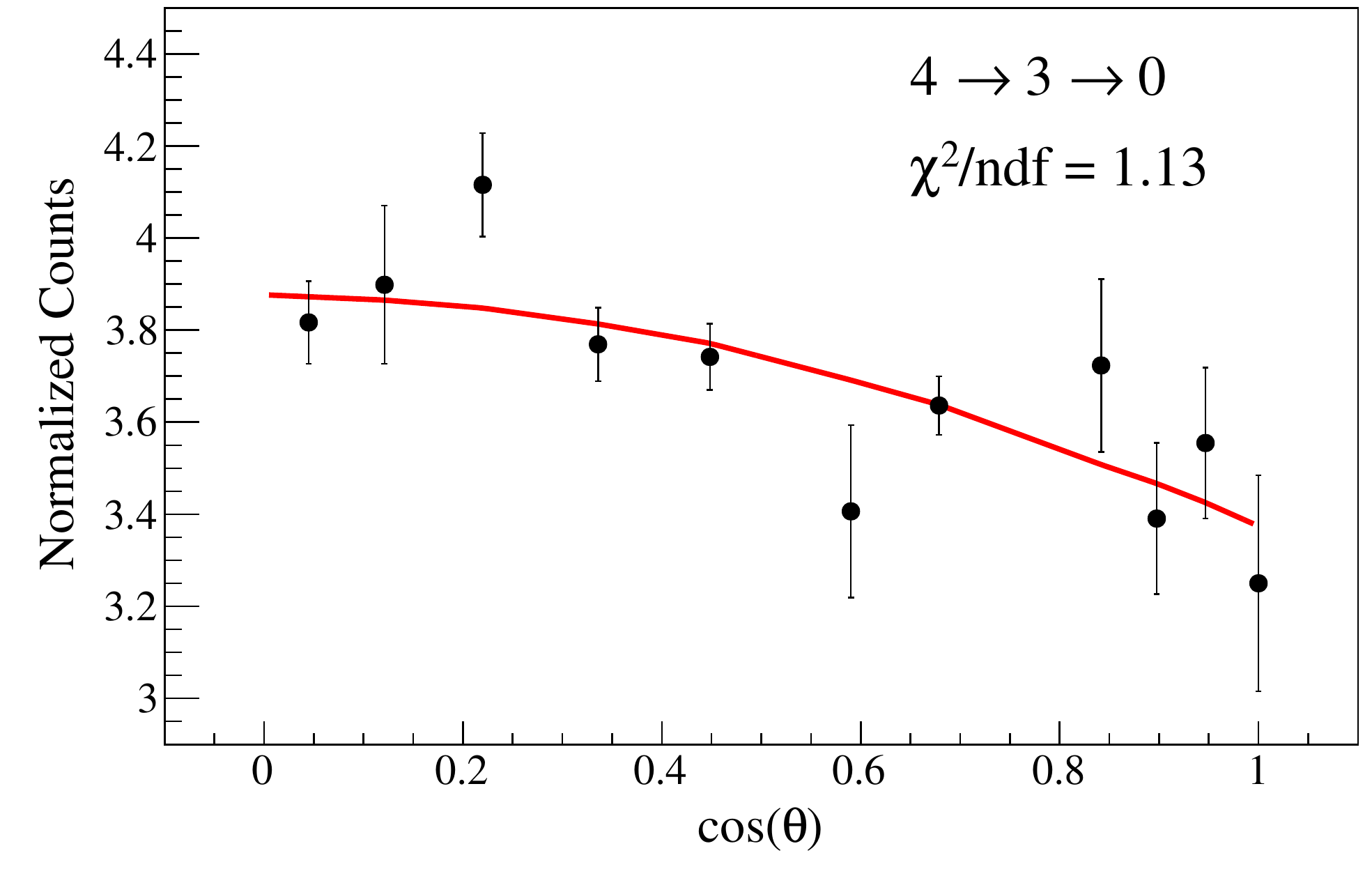}
\caption{\label{fig:ac}
The best-fit angular correlation for the 479-4352~keV $4^{-}\rightarrow3^{-}\rightarrow0^{+}$ cascade in $^{132}$Sn.
The experimental data points are shown in black, and the simulated curve is shown in red.
The fit corresponds to an $E2$/$M1$ mixing ratio of 0.02(2) for the 479-keV $4^{-}\rightarrow3^{-}$ transition.
}
\end{figure}

In order to fit the experimental angular correlations to Eq.~\ref{eq1}, it was necessary to take into account the finite size effects of GRIFFIN which tend to attenuate the $a_{2}$ and $a_{4}$ coefficients.
Therefore, simulations were performed according to Method 2 of Ref.~\cite{Smith2019}.
The simulations were based on the Geant4 toolkit~\cite{Agostinelli2003} with an additional extension to the radioactive decay classes to reproduce physical angular correlations~\cite{Ashfield2017}.
For each cascade, three simulations were performed that match the following distributions:
\begin{align}
\mathcal{Z}_{0}(\theta)&=1 \\
\mathcal{Z}_{2}(\theta)&=1+P_{2}(\cos\theta) \\
\mathcal{Z}_{4}(\theta)&=1+P_{4}(\cos\theta).
\end{align}
$\mathcal{Z}_{2}$ and $\mathcal{Z}_{4}$ represent nearly pure Legendre polynomials, and are simulated individually so that the attenuation effects can be separated.
$\mathcal{Z}_{0}$ is a simulation with no angular correlation effects and is used to normalize the outputs of the $\mathcal{Z}_{2}$ and $\mathcal{Z}_{4}$ simulations.
A linear combination of these simulations was used to fit to the data:
\begin{equation}
\mathcal{Z}_{sum}=A_{00}[(1-a_{2}-a_{4})\mathcal{Z}_{0}+a_{2}\mathcal{Z}_{2}+a_{4}\mathcal{Z}_{4}] \label{eq5}
\end{equation}
The simulated data were grouped and folded in the same manner as the experimental data.
A sample angular correlation is presented in Figure~\ref{fig:ac}, which shows a fit to the 479-4352~keV $4^{-}\rightarrow3^{-}\rightarrow0^{+}$ cascade in $^{132}$Sn.

For many of the angular correlations described in Ref.~\cite{Smith2019}, a mixing ratio is determined for a particular transition by performing a series of fits and varying the mixing ratio.
By adjusting the mixing ratio of either $\gamma$-ray transition in a cascade, theoretical values for $a_{2}$ and $a_{4}$ can be calculated, which are input into Equation~\ref{eq5}.
Then the simulated curve is fit to the experimental data with $A_{00}$ as a free parameter, and the $\chi^{2}$ of the fits can be determined as a function of $\delta$.
The minimum $\chi^{2}$ then determines the best-fit mixing ratio.
This process was performed for the 479-4352~keV $4^{-}\rightarrow3^{-}\rightarrow0^{+}$ cascade shown in Figure~\ref{fig:ac}.
The 4352-keV transition is constrained by the $0^{+}$ ground state to be a pure $E3$ transition.
However, the 479-keV transition between the $4^{-}$ and $3^{-}$ states can have a mixed $E2/M1$ multipolarity.
Therefore, the points were fit by varying the mixing ratio of the 479-keV transition.
In the figure, the best-fit curve is shown along with the experimental data points, which corresponds to an $E2$/$M1$ mixing ratio of 0.02(2), consistent with a pure $M1$ transition.

Three other angular correlations were analyzed in the current work; these are listed along with the 479-4352-keV cascade in Table~\ref{tab:ac}.
In general, the procedure for cascades without pure transitions is complex because of the two free mixing ratios.
The correlations in Table~\ref{tab:ac} were approximated by looking at the $\chi^{2}$ assuming zero mixing for both transitions in the cascades.
Like the 4352-keV transition, the multipolarity of the 4041-keV transition was fixed by the $0^{+}$ ground state.
The 375-, 300-, and 2380-keV transitions are $E2$, and $M3$ mixing is not expected to occur.
Because the mixing ratio for the 479-keV transition determined from the $\chi^{2}$ minimization was consistent with 0, this value was also used in the fit of the 479-4352-keV cascade.
In all cases, the reduced $\chi^{2}$ values are reasonable, confirming the spin and parity assignments of these states.

\begin{table}[b]
\caption{\label{tab:ac}
A summary of the $\gamma$-ray angular correlations for select cascades in $^{132}$Sn.
Listed are the first and second $\gamma$-ray transitions $E_{\gamma 1}$ and $E_{\gamma 2}$, the spin cascade used to calculate coefficients, and the reduced $\chi^{2}$ resulting from the fit.
The fits were made with the mixing ratios of both transitions set to 0, as discussed in the text.
}
\begin{tabular*}{\columnwidth}{@{\extracolsep{\fill}}c c c S}
\hline \hline
{$E_{\gamma 1}$\,(keV)} & {$E_{\gamma 2}$\,(keV)} & {Cascade} & $\chi^{2}/\mathrm{ndf}$ \\
\hline
375  & 4041 & $4^{+}\rightarrow 2^{+}\rightarrow 0^{+}$ & 1.04 \\
300  & 375  & $6^{+}\rightarrow 4^{+}\rightarrow 2^{+}$ & 1.91 \\
479  & 4352 & $4^{-}\rightarrow 3^{-}\rightarrow 0^{+}$ & 1.29 \\
2380 & 479  & $6^{-}\rightarrow 4^{-}\rightarrow 3^{-}$ & 0.78 \\
\hline \hline
\end{tabular*}
\end{table}

\section{\label{sec:theory} Theoretical Considerations}

Predictions for the excitation spectrum in $^{132}$Sn have been made using the ab initio VS-IMSRG.
The VS-IMSRG~\cite{Tsukiyama2012,Bogner2014,Morris2015,Stroberg2016,Stroberg2017,Stroberg2019} provides a prescription to generate approximately unitary transformations to decouple first a desired core energy followed by an appropriate valence-space Hamiltonian from the larger Hilbert space.
Calculations began with the 1.8/2.0(EM) chiral interaction of Refs.~\cite{Hebeler2011,Simonis2016,Simonis2017}, within a harmonic-oscillator basis of 15 major shells (i.e., $e=2n+l \leqslant e_{\mathrm{max}}=14$).
This interaction has been found to successfully reproduce ground-state and excitation energies throughout the medium- to heavy-mass region, including dripline properties~\cite{Simonis2017,Morris2017,Holt2019}.
The interaction was transformed to the Hartree-Fock basis and capture effects of $3N$ forces among valence nucleons with the ensemble normal ordering described in Ref.~\cite{Stroberg2017}.
Finally, for storage requirements, a cut of $e_{1}+e_{2}+e_{3} \leqslant E_{\mathrm{3Max}}$ was imposed for $3N$ matrix elements.

The imsrg++ code~\cite{Stroberg2017b}, adopting the Magnus formulation of the IMSRG~\cite{Morris2015,Hergert2016}, was used to generate transformations to perform the above decouplings, where in the IMSRG(2) approximation, all induced operators are truncated at the two-body level.
For the region near $^{132}$Sn, a core of $^{92}$Ni was used, and a valence-space Hamiltonian was decoupled for the ($0g_{7/2}$, $1d_{5/2}$, $1d_{3/2}$, $2s_{1/2}$) proton orbits and ($1d_{3/2}$, $2s_{1/}2$, $0h_{11/2}$, $1f_{7/2}$, $2p_{3/2}$) neutron orbits.
Finally the resulting valence-space Hamiltonians were diagonalized with KSHELL shell-model code~\cite{Shimizu2019} to obtain ground- and excited-state energies.

For heavier systems near $^{132}$Sn, as mentioned earlier, achieving convergence with respect to $E_{{3}\mathrm{Max}}$ has been the primary bottleneck to fully exploring the heavy region of nuclei.
For example, in recent calculations near $^{132}$Sn~\cite{Lascar2017,Manea2020,Arthuis2020}, energies were clearly not converged even at the $E_{\mathrm{3Max}}=18$ level, which was the previous computational limit.
Very recently, however, significant advances in treatment of $3N$ matrix elements have been made which allow extension to at least $E_{\mathrm{3Max}}=28$~\cite{Miyagi}. 
Here the ground-state energy is converged to better than 5~MeV (i.e., 0.5\%), and excited states are converged to better than 100~keV with $E_{\mathrm{3Max}}=22$~\cite{Miyagi}.

\begin{figure}[t]
\includegraphics[width=\columnwidth]{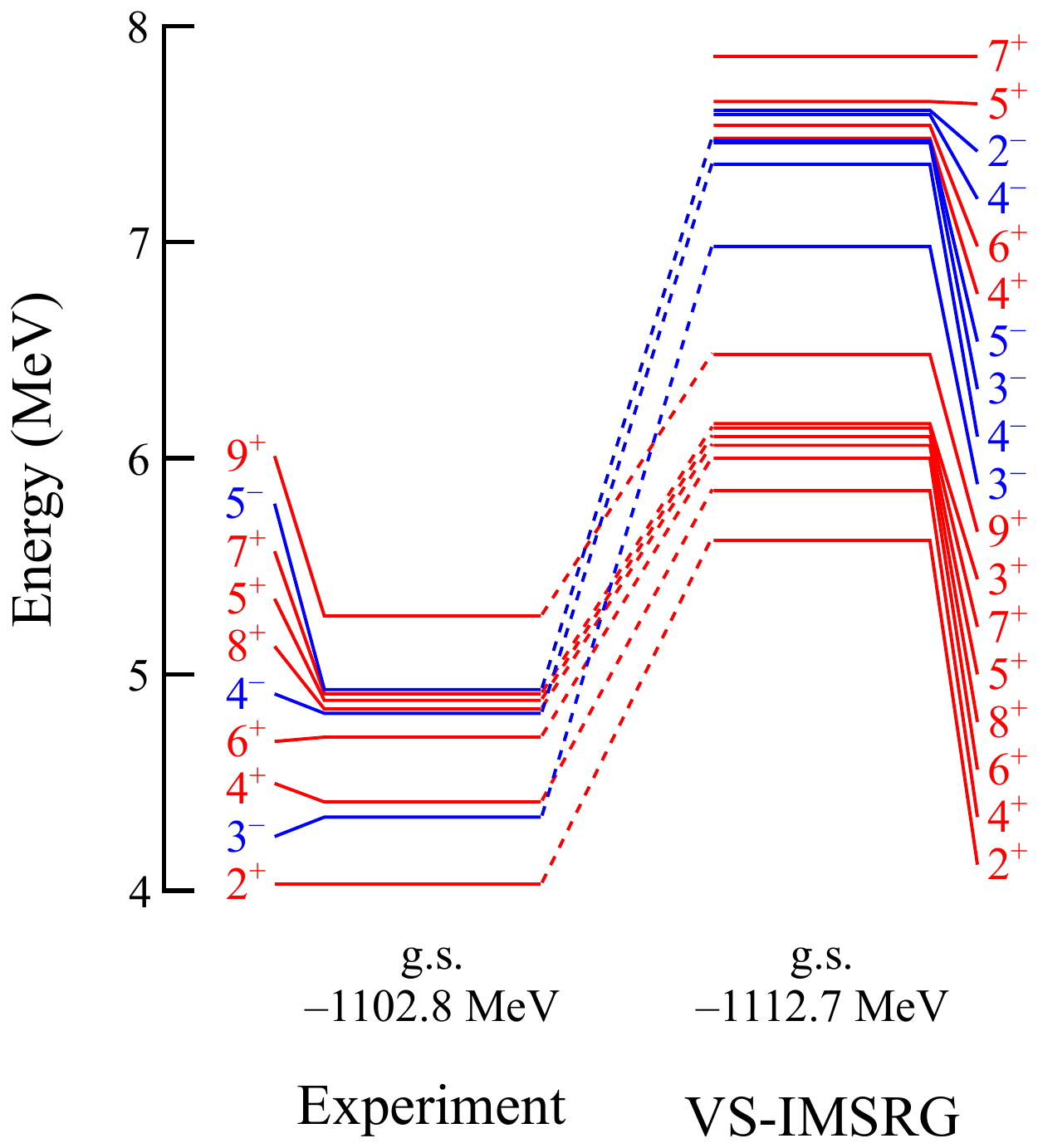}
\caption{\label{fig:theory}
A sample of the low-lying excited states in $^{132}$Sn from experiment (left) and calculated with the VS-IMSRG (right).
Positive-parity states are shown in red, and negative-parity states in blue.
The calculated levels are shown up to 8~MeV, and are connected to experimental levels by dashed lines where the levels can be directly matched.
The experimental and calculated ground-state (g.s.) energies are also listed below the graph.
}
\end{figure}

The results of these calculations are shown in Figure~\ref{fig:theory} alongside the experimental values.
The calculated ground-state energy agrees quite well with the experimental value, at the level of approximately 1\%.
The theoretical spectra is shown up to 8~MeV, and the experimental spectrum is shown for the positive-parity states up to the $9^{+}$ level at 5280~keV and for the negative parity states up to the $5^{-}$ level at 4942~keV.
These levels can be most directly compared to the theoretical spectrum, and are connected to their theoretical counterparts in the figure.
The lowest $1^{-}$ state is omitted from the figure, because it is highly sensitive to the center-of-mass correction discussed in recent cross-shell VS-IMSRG calculations~\cite{Miyagi2020}.
The energy of the measured and calculated ground state is also listed in the figure.
Comparing the spectra, there is a general $\sim1.5$-MeV shift for the positive-parity states, and a 2.5~MeV shift for the negative-parity states.
This trend is a common feature of ab initio calculations truncated at the two-body level as is done in the IMSRG(2) approximation~\cite{Miyagi2020,Taniuchi2019}, and significant improvement in this respect is expected when calculations are eventually advanced to the IMSRG(3) level.
Besides this shift, the ordering of positive-parity and negative-parity states are each individually well reproduced.
Additionally, the spacing between levels of the same parity is well matched, within 100~keV for almost all levels, indicating ab initio theory is well suited to describing energies in this region.

\section{\label{sec:conclusion} Conclusion}

The analysis of the $\beta$ and $\beta n$ decay of $^{132}$In using the GRIFFIN spectrometer is presented.
The half-life of the decay has been confirmed, and the $\beta$-delayed neutron emission probability of 12.3(4)\% has been measured for the first time with $\gamma$ rays.
Comparison of the $\gamma$ rays in $^{131}$Sn and $^{131}$Sb indicates that the $\beta n$ decay proceeds primarily through the high-spin $(11/2^{-})$ isomer in $^{131}$Sn.
Analysis of the $\gamma$ rays in $^{131}$Sb following the decay of the isomer allowed for several new spin and parity assignments. 
This characterization of the decay will also provide valuable input to astrophysical $r$-process calculations.
Within $^{132}$Sn, 70 $\gamma$-ray transitions among 29 excited states have been identified with precise $\beta$-feeding and $\log ft$ values calculated for the excited states, and these have been used to provide more restricted spin and parity assignments for several levels.
Angular correlations were able to confirm the spin assignments of a few of the strongly fed states.
Finally, VS-IMSRG calculations were presented which reproduce the spacing of several of the excited states.

\begin{acknowledgments}
We would like to thank the operations and beam delivery staff at TRIUMF for providing the radioactive beam. 
The first phase of the GRIFFIN infrastructure has been funded jointly by the Canada Foundation for Innovation, TRIUMF, and the University of Guelph.
TRIUMF receives federal funding via a contribution agreement through the National Research Council Canada (NRC).
This work was supported in part by the Natural Sciences and Engineering Research Council of Canada (NSERC).
The authors would like to thank J.~Simonis for providing the 1.8/2.0(EM) element files and S.R.~Stroberg for the imsrg++ code~\cite{Stroberg2017b} used to perform these calculations.
Computations were performed with an allocation of computing resources on Cedar at WestGrid and Compute Canada, and on the Oak Cluster at TRIUMF managed by the University of British Columbia department of Advanced Research Computing (ARC).
\end{acknowledgments}

\bibliography{Sn132_draft}

\begin{thebibliography}{66}%
\makeatletter
\providecommand \@ifxundefined [1]{%
 \@ifx{#1\undefined}
}%
\providecommand \@ifnum [1]{%
 \ifnum #1\expandafter \@firstoftwo
 \else \expandafter \@secondoftwo
 \fi
}%
\providecommand \@ifx [1]{%
 \ifx #1\expandafter \@firstoftwo
 \else \expandafter \@secondoftwo
 \fi
}%
\providecommand \natexlab [1]{#1}%
\providecommand \enquote  [1]{``#1''}%
\providecommand \bibnamefont  [1]{#1}%
\providecommand \bibfnamefont [1]{#1}%
\providecommand \citenamefont [1]{#1}%
\providecommand \href@noop [0]{\@secondoftwo}%
\providecommand \href [0]{\begingroup \@sanitize@url \@href}%
\providecommand \@href[1]{\@@startlink{#1}\@@href}%
\providecommand \@@href[1]{\endgroup#1\@@endlink}%
\providecommand \@sanitize@url [0]{\catcode `\\12\catcode `\$12\catcode
  `\&12\catcode `\#12\catcode `\^12\catcode `\_12\catcode `\%12\relax}%
\providecommand \@@startlink[1]{}%
\providecommand \@@endlink[0]{}%
\providecommand \url  [0]{\begingroup\@sanitize@url \@url }%
\providecommand \@url [1]{\endgroup\@href {#1}{\urlprefix }}%
\providecommand \urlprefix  [0]{URL }%
\providecommand \Eprint [0]{\href }%
\providecommand \doibase [0]{http://dx.doi.org/}%
\providecommand \selectlanguage [0]{\@gobble}%
\providecommand \bibinfo  [0]{\@secondoftwo}%
\providecommand \bibfield  [0]{\@secondoftwo}%
\providecommand \translation [1]{[#1]}%
\providecommand \BibitemOpen [0]{}%
\providecommand \bibitemStop [0]{}%
\providecommand \bibitemNoStop [0]{.\EOS\space}%
\providecommand \EOS [0]{\spacefactor3000\relax}%
\providecommand \BibitemShut  [1]{\csname bibitem#1\endcsname}%
\let\auto@bib@innerbib\@empty
\bibitem [{\citenamefont {Chartier}\ \emph {et~al.}(1996)\citenamefont
  {Chartier} \emph {et~al.}}]{Chartier1996}%
  \BibitemOpen
  \bibfield  {author} {\bibinfo {author} {\bibfnamefont {M.}~\bibnamefont
  {Chartier}} \emph {et~al.},\ }\href {\doibase 10.1103/PhysRevLett.77.2400}
  {\bibfield  {journal} {\bibinfo  {journal} {Phys. Rev. Lett.}\ }\textbf
  {\bibinfo {volume} {77}},\ \bibinfo {pages} {2400} (\bibinfo {year}
  {1996})}\BibitemShut {NoStop}%
\bibitem [{\citenamefont {Hinke}\ \emph {et~al.}(2012)\citenamefont {Hinke}
  \emph {et~al.}}]{Hinke2012}%
  \BibitemOpen
  \bibfield  {author} {\bibinfo {author} {\bibfnamefont {C.~B.}\ \bibnamefont
  {Hinke}} \emph {et~al.},\ }\href {\doibase 10.1038/nature11116} {\bibfield
  {journal} {\bibinfo  {journal} {Nature}\ }\textbf {\bibinfo {volume} {486}},\
  \bibinfo {pages} {341} (\bibinfo {year} {2012})}\BibitemShut {NoStop}%
\bibitem [{\citenamefont {Lubos}\ \emph {et~al.}(2019)\citenamefont {Lubos}
  \emph {et~al.}}]{Lubos2019}%
  \BibitemOpen
  \bibfield  {author} {\bibinfo {author} {\bibfnamefont {D.}~\bibnamefont
  {Lubos}} \emph {et~al.},\ }\href {\doibase 10.1103/PhysRevLett.122.222502}
  {\bibfield  {journal} {\bibinfo  {journal} {Phys. Rev. Lett.}\ }\textbf
  {\bibinfo {volume} {122}},\ \bibinfo {pages} {222502} (\bibinfo {year}
  {2019})}\BibitemShut {NoStop}%
\bibitem [{\citenamefont {Blomqvist}(1981)}]{Blomqvist1981}%
  \BibitemOpen
  \bibfield  {author} {\bibinfo {author} {\bibfnamefont {J.}~\bibnamefont
  {Blomqvist}},\ }in\ \href {\doibase 10.5170/CERN-1981-009.536} {\emph
  {\bibinfo {booktitle} {Proceedings of the 4th International Conference on
  Nuclei Far from Stability}}}\ (\bibinfo {year} {1981})\ pp.\ \bibinfo {pages}
  {536--541}\BibitemShut {NoStop}%
\bibitem [{\citenamefont {Fogelberg}\ \emph {et~al.}(1994)\citenamefont
  {Fogelberg}, \citenamefont {Hellstr\"{o}m}, \citenamefont {Jerrestam},
  \citenamefont {Mach}, \citenamefont {Blomqvist}, \citenamefont {Kerek},
  \citenamefont {Norlin},\ and\ \citenamefont {Omtvedt}}]{Fogelberg1994}%
  \BibitemOpen
  \bibfield  {author} {\bibinfo {author} {\bibfnamefont {B.}~\bibnamefont
  {Fogelberg}}, \bibinfo {author} {\bibfnamefont {M.}~\bibnamefont
  {Hellstr\"{o}m}}, \bibinfo {author} {\bibfnamefont {D.}~\bibnamefont
  {Jerrestam}}, \bibinfo {author} {\bibfnamefont {H.}~\bibnamefont {Mach}},
  \bibinfo {author} {\bibfnamefont {J.}~\bibnamefont {Blomqvist}}, \bibinfo
  {author} {\bibfnamefont {A.}~\bibnamefont {Kerek}}, \bibinfo {author}
  {\bibfnamefont {L.~O.}\ \bibnamefont {Norlin}}, \ and\ \bibinfo {author}
  {\bibfnamefont {J.~P.}\ \bibnamefont {Omtvedt}},\ }\href {\doibase
  10.1103/PhysRevLett.73.2413} {\bibfield  {journal} {\bibinfo  {journal}
  {Phys. Rev. Lett.}\ }\textbf {\bibinfo {volume} {73}},\ \bibinfo {pages}
  {2413} (\bibinfo {year} {1994})}\BibitemShut {NoStop}%
\bibitem [{\citenamefont {Bj\"{o}rnstad}\ \emph {et~al.}(1986)\citenamefont
  {Bj\"{o}rnstad} \emph {et~al.}}]{Bjornstad1986}%
  \BibitemOpen
  \bibfield  {author} {\bibinfo {author} {\bibfnamefont {T.}~\bibnamefont
  {Bj\"{o}rnstad}} \emph {et~al.},\ }\href {\doibase
  10.1016/0375-9474(86)90447-1} {\bibfield  {journal} {\bibinfo  {journal}
  {Nucl. Phys. A}\ }\textbf {\bibinfo {volume} {453}},\ \bibinfo {pages} {463}
  (\bibinfo {year} {1986})}\BibitemShut {NoStop}%
\bibitem [{\citenamefont {Fogelberg}\ \emph {et~al.}(1995)\citenamefont
  {Fogelberg}, \citenamefont {Hellstr\"{o}m}, \citenamefont {Jerrestam},
  \citenamefont {Mach}, \citenamefont {Blomqvist}, \citenamefont {Kerek},
  \citenamefont {Norlin},\ and\ \citenamefont {Omtvedt}}]{Fogelberg1995}%
  \BibitemOpen
  \bibfield  {author} {\bibinfo {author} {\bibfnamefont {B.}~\bibnamefont
  {Fogelberg}}, \bibinfo {author} {\bibfnamefont {M.}~\bibnamefont
  {Hellstr\"{o}m}}, \bibinfo {author} {\bibfnamefont {D.}~\bibnamefont
  {Jerrestam}}, \bibinfo {author} {\bibfnamefont {H.}~\bibnamefont {Mach}},
  \bibinfo {author} {\bibfnamefont {J.}~\bibnamefont {Blomqvist}}, \bibinfo
  {author} {\bibfnamefont {A.}~\bibnamefont {Kerek}}, \bibinfo {author}
  {\bibfnamefont {L.~O.}\ \bibnamefont {Norlin}}, \ and\ \bibinfo {author}
  {\bibfnamefont {J.~P.}\ \bibnamefont {Omtvedt}},\ }\href {\doibase
  10.1088/0031-8949/1995/t56/012} {\bibfield  {journal} {\bibinfo  {journal}
  {Phys. Scripta}\ }\textbf {\bibinfo {volume} {T56}},\ \bibinfo {pages} {79}
  (\bibinfo {year} {1995})}\BibitemShut {NoStop}%
\bibitem [{\citenamefont {Bhattacharyya}\ \emph {et~al.}(2001)\citenamefont
  {Bhattacharyya} \emph {et~al.}}]{Bhattacharyya2001}%
  \BibitemOpen
  \bibfield  {author} {\bibinfo {author} {\bibfnamefont {P.}~\bibnamefont
  {Bhattacharyya}} \emph {et~al.},\ }\href {\doibase
  10.1103/PhysRevLett.87.062502} {\bibfield  {journal} {\bibinfo  {journal}
  {Phys. Rev. Lett.}\ }\textbf {\bibinfo {volume} {87}},\ \bibinfo {pages}
  {062502} (\bibinfo {year} {2001})}\BibitemShut {NoStop}%
\bibitem [{\citenamefont {Radford}\ \emph {et~al.}(2004)\citenamefont {Radford}
  \emph {et~al.}}]{Radford2004}%
  \BibitemOpen
  \bibfield  {author} {\bibinfo {author} {\bibfnamefont {D.~C.}\ \bibnamefont
  {Radford}} \emph {et~al.},\ }\href {\doibase 10.1016/j.nuclphysa.2004.09.056}
  {\bibfield  {journal} {\bibinfo  {journal} {Nucl. Phys. A}\ }\textbf
  {\bibinfo {volume} {746}},\ \bibinfo {pages} {83c} (\bibinfo {year}
  {2004})}\BibitemShut {NoStop}%
\bibitem [{\citenamefont {Beene}\ \emph {et~al.}(2004)\citenamefont {Beene}
  \emph {et~al.}}]{Beene2004}%
  \BibitemOpen
  \bibfield  {author} {\bibinfo {author} {\bibfnamefont {J.~R.}\ \bibnamefont
  {Beene}} \emph {et~al.},\ }\href {\doibase 10.1016/j.nuclphysa.2004.09.143}
  {\bibfield  {journal} {\bibinfo  {journal} {Nucl. Phys. A}\ }\textbf
  {\bibinfo {volume} {746}},\ \bibinfo {pages} {471c} (\bibinfo {year}
  {2004})}\BibitemShut {NoStop}%
\bibitem [{\citenamefont {Radford}\ \emph {et~al.}(2005)\citenamefont {Radford}
  \emph {et~al.}}]{Radford2005}%
  \BibitemOpen
  \bibfield  {author} {\bibinfo {author} {\bibfnamefont {D.~C.}\ \bibnamefont
  {Radford}} \emph {et~al.},\ }\href {\doibase 10.1016/j.nuclphysa.2005.02.040}
  {\bibfield  {journal} {\bibinfo  {journal} {Nucl. Phys. A}\ }\textbf
  {\bibinfo {volume} {752}},\ \bibinfo {pages} {264c} (\bibinfo {year}
  {2005})}\BibitemShut {NoStop}%
\bibitem [{\citenamefont {Varner}\ \emph {et~al.}(2005)\citenamefont {Varner}
  \emph {et~al.}}]{Varner2005}%
  \BibitemOpen
  \bibfield  {author} {\bibinfo {author} {\bibfnamefont {R.~L.}\ \bibnamefont
  {Varner}} \emph {et~al.},\ }\href {\doibase 10.1140/epjad/i2005-06-128-7}
  {\bibfield  {journal} {\bibinfo  {journal} {Eur. Phys. J. A}\ }\textbf
  {\bibinfo {volume} {25}},\ \bibinfo {pages} {391} (\bibinfo {year}
  {2005})}\BibitemShut {NoStop}%
\bibitem [{\citenamefont {Rosiak}\ \emph {et~al.}(2018)\citenamefont {Rosiak}
  \emph {et~al.}}]{Rosiak2018}%
  \BibitemOpen
  \bibfield  {author} {\bibinfo {author} {\bibfnamefont {D.}~\bibnamefont
  {Rosiak}} \emph {et~al.},\ }\href {\doibase 10.1103/PhysRevLett.121.252501}
  {\bibfield  {journal} {\bibinfo  {journal} {Phys. Rev. Lett.}\ }\textbf
  {\bibinfo {volume} {121}},\ \bibinfo {pages} {252501} (\bibinfo {year}
  {2018})}\BibitemShut {NoStop}%
\bibitem [{\citenamefont {Piersa}\ \emph {et~al.}(2019)\citenamefont {Piersa}
  \emph {et~al.}}]{Piersa2019}%
  \BibitemOpen
  \bibfield  {author} {\bibinfo {author} {\bibfnamefont {M.}~\bibnamefont
  {Piersa}} \emph {et~al.},\ }\href {\doibase 10.1103/PhysRevC.99.024304}
  {\bibfield  {journal} {\bibinfo  {journal} {Phys. Rev. C}\ }\textbf {\bibinfo
  {volume} {99}},\ \bibinfo {pages} {024304} (\bibinfo {year}
  {2019})}\BibitemShut {NoStop}%
\bibitem [{\citenamefont {Benito}\ \emph {et~al.}(2020)\citenamefont {Benito}
  \emph {et~al.}}]{Benito2020}%
  \BibitemOpen
  \bibfield  {author} {\bibinfo {author} {\bibfnamefont {J.}~\bibnamefont
  {Benito}} \emph {et~al.},\ }\href@noop {} {\bibfield  {journal} {\bibinfo
  {journal} {Phys. Rev. C}\ } (\bibinfo {year} {2020})},\ \bibinfo {note} {(in
  press)}\BibitemShut {NoStop}%
\bibitem [{\citenamefont {Terasaki}\ \emph {et~al.}(2002)\citenamefont
  {Terasaki}, \citenamefont {Engel}, \citenamefont {Nazarewicz},\ and\
  \citenamefont {Stoitsov}}]{Terasaki2002}%
  \BibitemOpen
  \bibfield  {author} {\bibinfo {author} {\bibfnamefont {J.}~\bibnamefont
  {Terasaki}}, \bibinfo {author} {\bibfnamefont {J.}~\bibnamefont {Engel}},
  \bibinfo {author} {\bibfnamefont {W.}~\bibnamefont {Nazarewicz}}, \ and\
  \bibinfo {author} {\bibfnamefont {M.}~\bibnamefont {Stoitsov}},\ }\href
  {\doibase 10.1103/PhysRevC.66.054313} {\bibfield  {journal} {\bibinfo
  {journal} {Phys. Rev. C}\ }\textbf {\bibinfo {volume} {66}},\ \bibinfo
  {pages} {054313} (\bibinfo {year} {2002})}\BibitemShut {NoStop}%
\bibitem [{\citenamefont {Ansari}(2005)}]{Ansari2005}%
  \BibitemOpen
  \bibfield  {author} {\bibinfo {author} {\bibfnamefont {A.}~\bibnamefont
  {Ansari}},\ }\href {\doibase 10.1016/j.physletb.2005.07.031} {\bibfield
  {journal} {\bibinfo  {journal} {Phys. Lett. B}\ }\textbf {\bibinfo {volume}
  {623}},\ \bibinfo {pages} {37} (\bibinfo {year} {2005})}\BibitemShut
  {NoStop}%
\bibitem [{\citenamefont {Ansari}\ and\ \citenamefont
  {Ring}(2006)}]{Ansari2006}%
  \BibitemOpen
  \bibfield  {author} {\bibinfo {author} {\bibfnamefont {A.}~\bibnamefont
  {Ansari}}\ and\ \bibinfo {author} {\bibfnamefont {P.}~\bibnamefont {Ring}},\
  }\href {\doibase 10.1103/PhysRevC.74.054313} {\bibfield  {journal} {\bibinfo
  {journal} {Phys. Rev. C}\ }\textbf {\bibinfo {volume} {74}},\ \bibinfo
  {pages} {054313} (\bibinfo {year} {2006})}\BibitemShut {NoStop}%
\bibitem [{\citenamefont {Y\"{u}ksel}\ \emph {et~al.}(2018)\citenamefont
  {Y\"{u}ksel}, \citenamefont {Col\`{o}}, \citenamefont {Khan},\ and\
  \citenamefont {Niu}}]{Yuksel2018}%
  \BibitemOpen
  \bibfield  {author} {\bibinfo {author} {\bibfnamefont {E.}~\bibnamefont
  {Y\"{u}ksel}}, \bibinfo {author} {\bibfnamefont {G.}~\bibnamefont
  {Col\`{o}}}, \bibinfo {author} {\bibfnamefont {E.}~\bibnamefont {Khan}}, \
  and\ \bibinfo {author} {\bibfnamefont {Y.~F.}\ \bibnamefont {Niu}},\ }\href
  {\doibase 10.1103/PhysRevC.97.064308} {\bibfield  {journal} {\bibinfo
  {journal} {Phys. Rev. C}\ }\textbf {\bibinfo {volume} {97}},\ \bibinfo
  {pages} {064308} (\bibinfo {year} {2018})}\BibitemShut {NoStop}%
\bibitem [{\citenamefont {Mumpower}\ \emph {et~al.}(2016)\citenamefont
  {Mumpower}, \citenamefont {Surman}, \citenamefont {McLaughlin},\ and\
  \citenamefont {Aprahamian}}]{Mumpower2016}%
  \BibitemOpen
  \bibfield  {author} {\bibinfo {author} {\bibfnamefont {M.~R.}\ \bibnamefont
  {Mumpower}}, \bibinfo {author} {\bibfnamefont {R.}~\bibnamefont {Surman}},
  \bibinfo {author} {\bibfnamefont {G.~C.}\ \bibnamefont {McLaughlin}}, \ and\
  \bibinfo {author} {\bibfnamefont {A.}~\bibnamefont {Aprahamian}},\ }\href
  {\doibase 10.1016/j.ppnp.2015.09.001} {\bibfield  {journal} {\bibinfo
  {journal} {Progr. Part. Nucl. Physic}\ }\textbf {\bibinfo {volume} {86}},\
  \bibinfo {pages} {86} (\bibinfo {year} {2016})}\BibitemShut {NoStop}%
\bibitem [{\citenamefont {Abriola}\ \emph {et~al.}(2011)\citenamefont
  {Abriola}, \citenamefont {Singh},\ and\ \citenamefont
  {Dillmann}}]{Abriola2011}%
  \BibitemOpen
  \bibfield  {author} {\bibinfo {author} {\bibfnamefont {D.}~\bibnamefont
  {Abriola}}, \bibinfo {author} {\bibfnamefont {B.}~\bibnamefont {Singh}}, \
  and\ \bibinfo {author} {\bibfnamefont {I.}~\bibnamefont {Dillmann}},\
  }\href@noop {} {\bibfield  {journal} {\bibinfo  {journal} {IAEA Technical
  Report No. INDC(NDS)-0599}\ } (\bibinfo {year} {2011})}\BibitemShut {NoStop}%
\bibitem [{\citenamefont {Jungclaus}\ \emph {et~al.}(2016)\citenamefont
  {Jungclaus} \emph {et~al.}}]{Jungclaus2016}%
  \BibitemOpen
  \bibfield  {author} {\bibinfo {author} {\bibfnamefont {A.}~\bibnamefont
  {Jungclaus}} \emph {et~al.},\ }\href {\doibase 10.1103/PhysRevC.93.041301}
  {\bibfield  {journal} {\bibinfo  {journal} {Phys. Rev. C}\ }\textbf {\bibinfo
  {volume} {93}},\ \bibinfo {pages} {041301(R)} (\bibinfo {year}
  {2016})}\BibitemShut {NoStop}%
\bibitem [{\citenamefont {Dunlop}\ \emph {et~al.}(2019)\citenamefont {Dunlop}
  \emph {et~al.}}]{Dunlop2019}%
  \BibitemOpen
  \bibfield  {author} {\bibinfo {author} {\bibfnamefont {R.}~\bibnamefont
  {Dunlop}} \emph {et~al.},\ }\href {\doibase 10.1103/PhysRevC.99.045805}
  {\bibfield  {journal} {\bibinfo  {journal} {Phys. Rev. C}\ }\textbf {\bibinfo
  {volume} {99}},\ \bibinfo {pages} {045805} (\bibinfo {year}
  {2019})}\BibitemShut {NoStop}%
\bibitem [{\citenamefont {Lund}\ \emph {et~al.}(1980)\citenamefont {Lund},
  \citenamefont {Hoff}, \citenamefont {Aleklett}, \citenamefont {Glomset},\
  and\ \citenamefont {Rudstam}}]{Lund1980}%
  \BibitemOpen
  \bibfield  {author} {\bibinfo {author} {\bibfnamefont {E.}~\bibnamefont
  {Lund}}, \bibinfo {author} {\bibfnamefont {P.}~\bibnamefont {Hoff}}, \bibinfo
  {author} {\bibfnamefont {K.}~\bibnamefont {Aleklett}}, \bibinfo {author}
  {\bibfnamefont {O.}~\bibnamefont {Glomset}}, \ and\ \bibinfo {author}
  {\bibfnamefont {G.}~\bibnamefont {Rudstam}},\ }\href {\doibase
  10.1007/BF01438160} {\bibfield  {journal} {\bibinfo  {journal} {Z. Phys. A}\
  }\textbf {\bibinfo {volume} {294}},\ \bibinfo {pages} {233} (\bibinfo {year}
  {1980})}\BibitemShut {NoStop}%
\bibitem [{\citenamefont {Reeder}\ \emph {et~al.}(1986)\citenamefont {Reeder},
  \citenamefont {Warner}, \citenamefont {Edmiston}, \citenamefont {Gill},\ and\
  \citenamefont {Piotrowski}}]{Reeder1986}%
  \BibitemOpen
  \bibfield  {author} {\bibinfo {author} {\bibfnamefont {P.~L.}\ \bibnamefont
  {Reeder}}, \bibinfo {author} {\bibfnamefont {R.~A.}\ \bibnamefont {Warner}},
  \bibinfo {author} {\bibfnamefont {M.~D.}\ \bibnamefont {Edmiston}}, \bibinfo
  {author} {\bibfnamefont {R.~L.}\ \bibnamefont {Gill}}, \ and\ \bibinfo
  {author} {\bibfnamefont {A.}~\bibnamefont {Piotrowski}},\ }\enquote {\bibinfo
  {title} {{New Delayed-Neutron Precursors from TRISTAN}},}\ in\ \href
  {\doibase 10.1021/bk-1986-0324.ch025} {\emph {\bibinfo {booktitle} {Nuclei
  Off the Line of Stability}}}\ (\bibinfo {year} {1986})\ Chap.~\bibinfo
  {chapter} {25}, pp.\ \bibinfo {pages} {171--176}\BibitemShut {NoStop}%
\bibitem [{\citenamefont {Rudstam}\ \emph {et~al.}(1993)\citenamefont
  {Rudstam}, \citenamefont {Aleklett},\ and\ \citenamefont
  {Sihver}}]{Rudstam1993}%
  \BibitemOpen
  \bibfield  {author} {\bibinfo {author} {\bibfnamefont {G.}~\bibnamefont
  {Rudstam}}, \bibinfo {author} {\bibfnamefont {K.}~\bibnamefont {Aleklett}}, \
  and\ \bibinfo {author} {\bibfnamefont {L.}~\bibnamefont {Sihver}},\ }\href
  {\doibase 10.1006/adnd.1993.1001} {\bibfield  {journal} {\bibinfo  {journal}
  {Atom. Data Nucl. Data}\ }\textbf {\bibinfo {volume} {53}},\ \bibinfo {pages}
  {1} (\bibinfo {year} {1993})}\BibitemShut {NoStop}%
\bibitem [{\citenamefont {Morris}\ \emph {et~al.}(2018)\citenamefont {Morris},
  \citenamefont {Simonis}, \citenamefont {Stroberg}, \citenamefont {Stumpf},
  \citenamefont {Hagen}, \citenamefont {Holt}, \citenamefont {Jansen},
  \citenamefont {Papenbrock}, \citenamefont {Roth},\ and\ \citenamefont
  {Schwenk}}]{Morris2017}%
  \BibitemOpen
  \bibfield  {author} {\bibinfo {author} {\bibfnamefont {T.~D.}\ \bibnamefont
  {Morris}}, \bibinfo {author} {\bibfnamefont {J.}~\bibnamefont {Simonis}},
  \bibinfo {author} {\bibfnamefont {S.~R.}\ \bibnamefont {Stroberg}}, \bibinfo
  {author} {\bibfnamefont {C.}~\bibnamefont {Stumpf}}, \bibinfo {author}
  {\bibfnamefont {G.}~\bibnamefont {Hagen}}, \bibinfo {author} {\bibfnamefont
  {J.~D.}\ \bibnamefont {Holt}}, \bibinfo {author} {\bibfnamefont {G.~R.}\
  \bibnamefont {Jansen}}, \bibinfo {author} {\bibfnamefont {T.}~\bibnamefont
  {Papenbrock}}, \bibinfo {author} {\bibfnamefont {R.}~\bibnamefont {Roth}}, \
  and\ \bibinfo {author} {\bibfnamefont {A.}~\bibnamefont {Schwenk}},\ }\href
  {\doibase 10.1103/PhysRevLett.120.152503} {\bibfield  {journal} {\bibinfo
  {journal} {Phys. Rev. Lett.}\ }\textbf {\bibinfo {volume} {120}},\ \bibinfo
  {pages} {152503} (\bibinfo {year} {2018})}\BibitemShut {NoStop}%
\bibitem [{\citenamefont {Gysbers}\ \emph {et~al.}(2019)\citenamefont {Gysbers}
  \emph {et~al.}}]{Gysbers2019}%
  \BibitemOpen
  \bibfield  {author} {\bibinfo {author} {\bibfnamefont {P.}~\bibnamefont
  {Gysbers}} \emph {et~al.},\ }\href {\doibase 10.1038/s41567-019-0450-7}
  {\bibfield  {journal} {\bibinfo  {journal} {Nature Phys.}\ }\textbf {\bibinfo
  {volume} {15}},\ \bibinfo {pages} {428} (\bibinfo {year} {2019})}\BibitemShut
  {NoStop}%
\bibitem [{\citenamefont {Lascar}\ \emph {et~al.}(2017)\citenamefont {Lascar}
  \emph {et~al.}}]{Lascar2017}%
  \BibitemOpen
  \bibfield  {author} {\bibinfo {author} {\bibfnamefont {D.}~\bibnamefont
  {Lascar}} \emph {et~al.},\ }\href {\doibase 10.1103/PhysRevC.96.044323}
  {\bibfield  {journal} {\bibinfo  {journal} {Phys. Rev. C}\ }\textbf {\bibinfo
  {volume} {96}},\ \bibinfo {pages} {044323} (\bibinfo {year}
  {2017})}\BibitemShut {NoStop}%
\bibitem [{\citenamefont {Manea}\ \emph {et~al.}(2020)\citenamefont {Manea}
  \emph {et~al.}}]{Manea2020}%
  \BibitemOpen
  \bibfield  {author} {\bibinfo {author} {\bibfnamefont {V.}~\bibnamefont
  {Manea}} \emph {et~al.},\ }\href {\doibase 10.1103/PhysRevLett.124.092502}
  {\bibfield  {journal} {\bibinfo  {journal} {Phys. Rev. Lett.}\ }\textbf
  {\bibinfo {volume} {124}},\ \bibinfo {pages} {092502} (\bibinfo {year}
  {2020})}\BibitemShut {NoStop}%
\bibitem [{\citenamefont {Dilling}\ \emph {et~al.}(2014)\citenamefont
  {Dilling}, \citenamefont {Kr\"{u}cken},\ and\ \citenamefont
  {Ball}}]{Dilling2014}%
  \BibitemOpen
  \bibfield  {author} {\bibinfo {author} {\bibfnamefont {J.}~\bibnamefont
  {Dilling}}, \bibinfo {author} {\bibfnamefont {R.}~\bibnamefont
  {Kr\"{u}cken}}, \ and\ \bibinfo {author} {\bibfnamefont {G.}~\bibnamefont
  {Ball}},\ }\href {\doibase 10.1007/s10751-013-0877-7} {\bibfield  {journal}
  {\bibinfo  {journal} {Hyperfine Interact.}\ }\textbf {\bibinfo {volume}
  {225}},\ \bibinfo {pages} {1} (\bibinfo {year} {2014})}\BibitemShut {NoStop}%
\bibitem [{\citenamefont {Raeder}\ \emph {et~al.}(2014)\citenamefont {Raeder},
  \citenamefont {Heggen}, \citenamefont {Lassen}, \citenamefont {Ames},
  \citenamefont {Bishop}, \citenamefont {Bricault}, \citenamefont {Kunz},
  \citenamefont {Mj{\o}s},\ and\ \citenamefont {Teigelh\"{o}fer}}]{Raeder2014}%
  \BibitemOpen
  \bibfield  {author} {\bibinfo {author} {\bibfnamefont {S.}~\bibnamefont
  {Raeder}}, \bibinfo {author} {\bibfnamefont {H.}~\bibnamefont {Heggen}},
  \bibinfo {author} {\bibfnamefont {J.}~\bibnamefont {Lassen}}, \bibinfo
  {author} {\bibfnamefont {F.}~\bibnamefont {Ames}}, \bibinfo {author}
  {\bibfnamefont {D.}~\bibnamefont {Bishop}}, \bibinfo {author} {\bibfnamefont
  {P.}~\bibnamefont {Bricault}}, \bibinfo {author} {\bibfnamefont
  {P.}~\bibnamefont {Kunz}}, \bibinfo {author} {\bibfnamefont {A.}~\bibnamefont
  {Mj{\o}s}}, \ and\ \bibinfo {author} {\bibnamefont {Teigelh\"{o}fer}},\
  }\href {\doibase 10.1063/1.4868496} {\bibfield  {journal} {\bibinfo
  {journal} {Rev. Sci. Instrum.}\ }\textbf {\bibinfo {volume} {85}},\ \bibinfo
  {pages} {033309} (\bibinfo {year} {2014})}\BibitemShut {NoStop}%
\bibitem [{\citenamefont {Svensson}\ and\ \citenamefont
  {Garnsworthy}(2014)}]{Svensson2014}%
  \BibitemOpen
  \bibfield  {author} {\bibinfo {author} {\bibfnamefont {C.~E.}\ \bibnamefont
  {Svensson}}\ and\ \bibinfo {author} {\bibfnamefont {A.~B.}\ \bibnamefont
  {Garnsworthy}},\ }\href {\doibase 10.1007/s10751-013-0889-3} {\bibfield
  {journal} {\bibinfo  {journal} {Hyperfine Interact.}\ }\textbf {\bibinfo
  {volume} {225}},\ \bibinfo {pages} {127} (\bibinfo {year}
  {2014})}\BibitemShut {NoStop}%
\bibitem [{\citenamefont {Garnsworthy}\ \emph {et~al.}(2019)\citenamefont
  {Garnsworthy} \emph {et~al.}}]{Garnsworthy2019}%
  \BibitemOpen
  \bibfield  {author} {\bibinfo {author} {\bibfnamefont {A.~B.}\ \bibnamefont
  {Garnsworthy}} \emph {et~al.},\ }\href {\doibase 10.1016/j.nima.2018.11.115}
  {\bibfield  {journal} {\bibinfo  {journal} {Nucl. Instrum. Meth. Phys. Res.
  A}\ }\textbf {\bibinfo {volume} {918}},\ \bibinfo {pages} {9} (\bibinfo
  {year} {2019})}\BibitemShut {NoStop}%
\bibitem [{\citenamefont {Garnsworthy}\ \emph {et~al.}(2017)\citenamefont
  {Garnsworthy} \emph {et~al.}}]{Garnsworthy2017}%
  \BibitemOpen
  \bibfield  {author} {\bibinfo {author} {\bibfnamefont {A.~B.}\ \bibnamefont
  {Garnsworthy}} \emph {et~al.},\ }\href {\doibase 10.1016/j.nima.2017.02.040}
  {\bibfield  {journal} {\bibinfo  {journal} {Nucl. Instrum. Meth. Phys. Res.
  A}\ }\textbf {\bibinfo {volume} {853}},\ \bibinfo {pages} {85} (\bibinfo
  {year} {2017})}\BibitemShut {NoStop}%
\bibitem [{\citenamefont {Rizwan}\ \emph {et~al.}(2016)\citenamefont {Rizwan}
  \emph {et~al.}}]{Rizwan2016}%
  \BibitemOpen
  \bibfield  {author} {\bibinfo {author} {\bibfnamefont {U.}~\bibnamefont
  {Rizwan}} \emph {et~al.},\ }\href {\doibase 10.1016/j.nima.2016.03.016}
  {\bibfield  {journal} {\bibinfo  {journal} {Nucl. Instrum. Meth. Phys. Res.
  A}\ }\textbf {\bibinfo {volume} {820}},\ \bibinfo {pages} {126} (\bibinfo
  {year} {2016})}\BibitemShut {NoStop}%
\bibitem [{\citenamefont {Wang}\ \emph {et~al.}(2017)\citenamefont {Wang},
  \citenamefont {Gaudi}, \citenamefont {Kondev}, \citenamefont {Huang},
  \citenamefont {Naimi},\ and\ \citenamefont {Xu}}]{Wang2017}%
  \BibitemOpen
  \bibfield  {author} {\bibinfo {author} {\bibfnamefont {M.}~\bibnamefont
  {Wang}}, \bibinfo {author} {\bibfnamefont {G.}~\bibnamefont {Gaudi}},
  \bibinfo {author} {\bibfnamefont {F.~G.}\ \bibnamefont {Kondev}}, \bibinfo
  {author} {\bibfnamefont {W.~J.}\ \bibnamefont {Huang}}, \bibinfo {author}
  {\bibfnamefont {S.}~\bibnamefont {Naimi}}, \ and\ \bibinfo {author}
  {\bibfnamefont {X.}~\bibnamefont {Xu}},\ }\href {\doibase
  10.1088/1674-1137/41/3/030003} {\bibfield  {journal} {\bibinfo  {journal}
  {Chinese Phys. C}\ }\textbf {\bibinfo {volume} {41}},\ \bibinfo {pages}
  {030003} (\bibinfo {year} {2017})}\BibitemShut {NoStop}%
\bibitem [{\citenamefont {Singh}(2018{\natexlab{a}})}]{Singh132In}%
  \BibitemOpen
  \bibfield  {author} {\bibinfo {author} {\bibfnamefont {B.}~\bibnamefont
  {Singh}},\ }\href@noop {} {\enquote {\bibinfo {title} {Nuclear data sheets
  for $^{132}\mathrm{In}$},}\ } (\bibinfo {year}
  {2018}{\natexlab{a}})\BibitemShut {NoStop}%
\bibitem [{\citenamefont {Khazov}\ \emph {et~al.}(2005)\citenamefont {Khazov},
  \citenamefont {Rodionov}, \citenamefont {Sakharov},\ and\ \citenamefont
  {Singh}}]{Khazov2005}%
  \BibitemOpen
  \bibfield  {author} {\bibinfo {author} {\bibfnamefont {Y.}~\bibnamefont
  {Khazov}}, \bibinfo {author} {\bibfnamefont {A.~A.}\ \bibnamefont
  {Rodionov}}, \bibinfo {author} {\bibfnamefont {S.}~\bibnamefont {Sakharov}},
  \ and\ \bibinfo {author} {\bibfnamefont {B.}~\bibnamefont {Singh}},\ }\href
  {\doibase 10.1016/j.nds.2005.03.001} {\bibfield  {journal} {\bibinfo
  {journal} {Nucl. Data Sheets}\ }\textbf {\bibinfo {volume} {104}},\ \bibinfo
  {pages} {497} (\bibinfo {year} {2005})}\BibitemShut {NoStop}%
\bibitem [{\citenamefont {Kib\'{e}di}\ \emph {et~al.}(2008)\citenamefont
  {Kib\'{e}di}, \citenamefont {Burrows}, \citenamefont {Trzhaskovskaya},
  \citenamefont {Davidson},\ and\ \citenamefont {Nestor~Jr.}}]{Kibedi2008}%
  \BibitemOpen
  \bibfield  {author} {\bibinfo {author} {\bibfnamefont {T.}~\bibnamefont
  {Kib\'{e}di}}, \bibinfo {author} {\bibfnamefont {T.~W.}\ \bibnamefont
  {Burrows}}, \bibinfo {author} {\bibfnamefont {M.~B.}\ \bibnamefont
  {Trzhaskovskaya}}, \bibinfo {author} {\bibfnamefont {P.~M.}\ \bibnamefont
  {Davidson}}, \ and\ \bibinfo {author} {\bibfnamefont {C.~W.}\ \bibnamefont
  {Nestor~Jr.}},\ }\href {\doibase 10.1016/j.nima.2008.02.051} {\bibfield
  {journal} {\bibinfo  {journal} {Nucl. Instrum. Meth. Phys. Res. A}\ }\textbf
  {\bibinfo {volume} {589}},\ \bibinfo {pages} {202} (\bibinfo {year}
  {2008})}\BibitemShut {NoStop}%
\bibitem [{\citenamefont {Khazov}\ \emph {et~al.}(2006)\citenamefont {Khazov},
  \citenamefont {Mitropolsky},\ and\ \citenamefont {Rodionov}}]{Khazov2006}%
  \BibitemOpen
  \bibfield  {author} {\bibinfo {author} {\bibfnamefont {Y.}~\bibnamefont
  {Khazov}}, \bibinfo {author} {\bibfnamefont {I.}~\bibnamefont {Mitropolsky}},
  \ and\ \bibinfo {author} {\bibfnamefont {A.}~\bibnamefont {Rodionov}},\
  }\href {\doibase 10.1016/j.nds.2006.10.001} {\bibfield  {journal} {\bibinfo
  {journal} {Nucl. Data Sheets}\ }\textbf {\bibinfo {volume} {107}},\ \bibinfo
  {pages} {2715} (\bibinfo {year} {2006})}\BibitemShut {NoStop}%
\bibitem [{\citenamefont {Stone}\ \emph {et~al.}(1988)\citenamefont {Stone},
  \citenamefont {Faller}, \citenamefont {Robertson},\ and\ \citenamefont
  {Walters}}]{Stone1988}%
  \BibitemOpen
  \bibfield  {author} {\bibinfo {author} {\bibfnamefont {C.~A.}\ \bibnamefont
  {Stone}}, \bibinfo {author} {\bibfnamefont {S.~H.}\ \bibnamefont {Faller}},
  \bibinfo {author} {\bibfnamefont {J.~D.}\ \bibnamefont {Robertson}}, \ and\
  \bibinfo {author} {\bibfnamefont {W.~B.}\ \bibnamefont {Walters}},\ }\href
  {\doibase 10.1063/1.37013} {\bibfield  {journal} {\bibinfo  {journal} {AIP
  Conf. Proc.}\ }\textbf {\bibinfo {volume} {164}},\ \bibinfo {pages} {429}
  (\bibinfo {year} {1988})}\BibitemShut {NoStop}%
\bibitem [{\citenamefont {Fogelberg}\ \emph {et~al.}(2004)\citenamefont
  {Fogelberg} \emph {et~al.}}]{Fogelberg2004}%
  \BibitemOpen
  \bibfield  {author} {\bibinfo {author} {\bibfnamefont {B.}~\bibnamefont
  {Fogelberg}} \emph {et~al.},\ }\href {\doibase 10.1103/PhysRevC.70.034312}
  {\bibfield  {journal} {\bibinfo  {journal} {Phys. Rev. C}\ }\textbf {\bibinfo
  {volume} {70}},\ \bibinfo {pages} {034312} (\bibinfo {year}
  {2004})}\BibitemShut {NoStop}%
\bibitem [{\citenamefont {Singh}(2018{\natexlab{b}})}]{Singh132Sn}%
  \BibitemOpen
  \bibfield  {author} {\bibinfo {author} {\bibfnamefont {B.}~\bibnamefont
  {Singh}},\ }\href@noop {} {\enquote {\bibinfo {title} {Nuclear data sheets
  for $^{132}\mathrm{Sn}$},}\ } (\bibinfo {year}
  {2018}{\natexlab{b}})\BibitemShut {NoStop}%
\bibitem [{\citenamefont {Smith}\ \emph {et~al.}(2019)\citenamefont {Smith},
  \citenamefont {MacLean}, \citenamefont {Ashfield}, \citenamefont {Chester},
  \citenamefont {Garnsworthy},\ and\ \citenamefont {Svensson}}]{Smith2019}%
  \BibitemOpen
  \bibfield  {author} {\bibinfo {author} {\bibfnamefont {J.~K.}\ \bibnamefont
  {Smith}}, \bibinfo {author} {\bibfnamefont {A.~D.}\ \bibnamefont {MacLean}},
  \bibinfo {author} {\bibfnamefont {W.}~\bibnamefont {Ashfield}}, \bibinfo
  {author} {\bibfnamefont {A.}~\bibnamefont {Chester}}, \bibinfo {author}
  {\bibfnamefont {A.~B.}\ \bibnamefont {Garnsworthy}}, \ and\ \bibinfo {author}
  {\bibfnamefont {C.~E.}\ \bibnamefont {Svensson}},\ }\href {\doibase
  10.1016/j.nima.2018.10.097} {\bibfield  {journal} {\bibinfo  {journal} {Nucl.
  Instrum. Meth. Phys. Res. A}\ }\textbf {\bibinfo {volume} {922}},\ \bibinfo
  {pages} {47} (\bibinfo {year} {2019})}\BibitemShut {NoStop}%
\bibitem [{\citenamefont {Ortner}\ \emph {et~al.}(2020)\citenamefont {Ortner}
  \emph {et~al.}}]{Ortner2020}%
  \BibitemOpen
  \bibfield  {author} {\bibinfo {author} {\bibfnamefont {K.}~\bibnamefont
  {Ortner}} \emph {et~al.},\ }\href@noop {} {\bibfield  {journal} {\bibinfo
  {journal} {Phys. Rev. C}\ } (\bibinfo {year} {2020})},\ \bibinfo {note} {(in
  press)}\BibitemShut {NoStop}%
\bibitem [{\citenamefont {Pore}\ \emph {et~al.}(2019)\citenamefont {Pore} \emph
  {et~al.}}]{Pore2019}%
  \BibitemOpen
  \bibfield  {author} {\bibinfo {author} {\bibfnamefont {J.~L.}\ \bibnamefont
  {Pore}} \emph {et~al.},\ }\href {\doibase 10.1103/PhysRevC.100.054327}
  {\bibfield  {journal} {\bibinfo  {journal} {Phys. Rev. C}\ }\textbf {\bibinfo
  {volume} {100}},\ \bibinfo {pages} {054327} (\bibinfo {year}
  {2019})}\BibitemShut {NoStop}%
\bibitem [{\citenamefont {Agostinelli}\ \emph {et~al.}(2003)\citenamefont
  {Agostinelli} \emph {et~al.}}]{Agostinelli2003}%
  \BibitemOpen
  \bibfield  {author} {\bibinfo {author} {\bibfnamefont {S.}~\bibnamefont
  {Agostinelli}} \emph {et~al.},\ }\href {\doibase
  10.1016/S0168-9002(03)01368-8} {\bibfield  {journal} {\bibinfo  {journal}
  {Nucl. Instrum. Meth. Phys. Res. A}\ }\textbf {\bibinfo {volume} {506}},\
  \bibinfo {pages} {250} (\bibinfo {year} {2003})}\BibitemShut {NoStop}%
\bibitem [{\citenamefont {Ashfield}\ \emph {et~al.}(2017)\citenamefont
  {Ashfield}, \citenamefont {Rand}, \citenamefont {Bildstein},\ and\
  \citenamefont {Smith}}]{Ashfield2017}%
  \BibitemOpen
  \bibfield  {author} {\bibinfo {author} {\bibfnamefont {W.~H.}\ \bibnamefont
  {Ashfield}}, \bibinfo {author} {\bibfnamefont {E.~T.}\ \bibnamefont {Rand}},
  \bibinfo {author} {\bibfnamefont {V.}~\bibnamefont {Bildstein}}, \ and\
  \bibinfo {author} {\bibfnamefont {J.~K.}\ \bibnamefont {Smith}},\ }\href
  {\doibase 10.5281/zenodo.1120421} {\enquote {\bibinfo {title} {{Geant4 Gamma
  Gamma Angular Correlations}},}\ } (\bibinfo {year} {2017})\BibitemShut
  {NoStop}%
\bibitem [{\citenamefont {Tsukiyama}\ \emph {et~al.}(2012)\citenamefont
  {Tsukiyama}, \citenamefont {Bogner},\ and\ \citenamefont
  {Schwenk}}]{Tsukiyama2012}%
  \BibitemOpen
  \bibfield  {author} {\bibinfo {author} {\bibfnamefont {K.}~\bibnamefont
  {Tsukiyama}}, \bibinfo {author} {\bibfnamefont {S.~K.}\ \bibnamefont
  {Bogner}}, \ and\ \bibinfo {author} {\bibfnamefont {A.}~\bibnamefont
  {Schwenk}},\ }\href {\doibase 10.1103/PhysRevC.85.061304} {\bibfield
  {journal} {\bibinfo  {journal} {Phys. Rev. C}\ }\textbf {\bibinfo {volume}
  {85}},\ \bibinfo {pages} {061304(R)} (\bibinfo {year} {2012})}\BibitemShut
  {NoStop}%
\bibitem [{\citenamefont {Bogner}\ \emph {et~al.}(2014)\citenamefont {Bogner},
  \citenamefont {Hergert}, \citenamefont {Holt}, \citenamefont {Schwenk},
  \citenamefont {Binder}, \citenamefont {Calci}, \citenamefont {Langhammer},\
  and\ \citenamefont {Roth}}]{Bogner2014}%
  \BibitemOpen
  \bibfield  {author} {\bibinfo {author} {\bibfnamefont {S.~K.}\ \bibnamefont
  {Bogner}}, \bibinfo {author} {\bibfnamefont {H.}~\bibnamefont {Hergert}},
  \bibinfo {author} {\bibfnamefont {J.~D.}\ \bibnamefont {Holt}}, \bibinfo
  {author} {\bibfnamefont {A.}~\bibnamefont {Schwenk}}, \bibinfo {author}
  {\bibfnamefont {S.}~\bibnamefont {Binder}}, \bibinfo {author} {\bibfnamefont
  {A.}~\bibnamefont {Calci}}, \bibinfo {author} {\bibfnamefont
  {J.}~\bibnamefont {Langhammer}}, \ and\ \bibinfo {author} {\bibfnamefont
  {R.}~\bibnamefont {Roth}},\ }\href {\doibase 10.1103/PhysRevLett.113.142501}
  {\bibfield  {journal} {\bibinfo  {journal} {Phys. Rev. Lett.}\ }\textbf
  {\bibinfo {volume} {113}},\ \bibinfo {pages} {142501} (\bibinfo {year}
  {2014})}\BibitemShut {NoStop}%
\bibitem [{\citenamefont {Morris}\ \emph {et~al.}(2015)\citenamefont {Morris},
  \citenamefont {Parzuchowski},\ and\ \citenamefont {Bogner}}]{Morris2015}%
  \BibitemOpen
  \bibfield  {author} {\bibinfo {author} {\bibfnamefont {T.~D.}\ \bibnamefont
  {Morris}}, \bibinfo {author} {\bibfnamefont {N.~M.}\ \bibnamefont
  {Parzuchowski}}, \ and\ \bibinfo {author} {\bibfnamefont {S.~K.}\
  \bibnamefont {Bogner}},\ }\href {\doibase 10.1103/PhysRevC.92.034331}
  {\bibfield  {journal} {\bibinfo  {journal} {Phys. Rev. C}\ }\textbf {\bibinfo
  {volume} {92}},\ \bibinfo {pages} {034331} (\bibinfo {year}
  {2015})}\BibitemShut {NoStop}%
\bibitem [{\citenamefont {Stroberg}\ \emph {et~al.}(2016)\citenamefont
  {Stroberg}, \citenamefont {Hergert}, \citenamefont {Holt}, \citenamefont
  {Bogner},\ and\ \citenamefont {Schwenk}}]{Stroberg2016}%
  \BibitemOpen
  \bibfield  {author} {\bibinfo {author} {\bibfnamefont {S.~R.}\ \bibnamefont
  {Stroberg}}, \bibinfo {author} {\bibfnamefont {H.}~\bibnamefont {Hergert}},
  \bibinfo {author} {\bibfnamefont {J.~D.}\ \bibnamefont {Holt}}, \bibinfo
  {author} {\bibfnamefont {S.~K.}\ \bibnamefont {Bogner}}, \ and\ \bibinfo
  {author} {\bibfnamefont {A.}~\bibnamefont {Schwenk}},\ }\href {\doibase
  10.1103/PhysRevC.93.051301} {\bibfield  {journal} {\bibinfo  {journal} {Phys.
  Rev. C}\ }\textbf {\bibinfo {volume} {93}},\ \bibinfo {pages} {051301(R)}
  (\bibinfo {year} {2016})}\BibitemShut {NoStop}%
\bibitem [{\citenamefont {Stroberg}\ \emph {et~al.}(2017)\citenamefont
  {Stroberg}, \citenamefont {Calci}, \citenamefont {Hergert}, \citenamefont
  {Holt}, \citenamefont {Bogner}, \citenamefont {Roth},\ and\ \citenamefont
  {Schwenk}}]{Stroberg2017}%
  \BibitemOpen
  \bibfield  {author} {\bibinfo {author} {\bibfnamefont {S.~R.}\ \bibnamefont
  {Stroberg}}, \bibinfo {author} {\bibfnamefont {A.}~\bibnamefont {Calci}},
  \bibinfo {author} {\bibfnamefont {H.}~\bibnamefont {Hergert}}, \bibinfo
  {author} {\bibfnamefont {J.~D.}\ \bibnamefont {Holt}}, \bibinfo {author}
  {\bibfnamefont {S.~K.}\ \bibnamefont {Bogner}}, \bibinfo {author}
  {\bibfnamefont {R.}~\bibnamefont {Roth}}, \ and\ \bibinfo {author}
  {\bibfnamefont {A.}~\bibnamefont {Schwenk}},\ }\href {\doibase
  10.1103/PhysRevLett.118.032502} {\bibfield  {journal} {\bibinfo  {journal}
  {Phys. Rev. Lett.}\ }\textbf {\bibinfo {volume} {118}},\ \bibinfo {pages}
  {032502} (\bibinfo {year} {2017})}\BibitemShut {NoStop}%
\bibitem [{\citenamefont {Stroberg}\ \emph {et~al.}(2019)\citenamefont
  {Stroberg}, \citenamefont {Bogner}, \citenamefont {Hergert},\ and\
  \citenamefont {Holt}}]{Stroberg2019}%
  \BibitemOpen
  \bibfield  {author} {\bibinfo {author} {\bibfnamefont {S.~R.}\ \bibnamefont
  {Stroberg}}, \bibinfo {author} {\bibfnamefont {S.~K.}\ \bibnamefont
  {Bogner}}, \bibinfo {author} {\bibfnamefont {H.}~\bibnamefont {Hergert}}, \
  and\ \bibinfo {author} {\bibfnamefont {J.~D.}\ \bibnamefont {Holt}},\ }\href
  {\doibase 10.1146/annurev-nucl-101917-021120} {\bibfield  {journal} {\bibinfo
   {journal} {Ann. Rev. Nucl. Part. Sci.}\ }\textbf {\bibinfo {volume} {69}},\
  \bibinfo {pages} {307} (\bibinfo {year} {2019})}\BibitemShut {NoStop}%
\bibitem [{\citenamefont {Hebeler}\ \emph {et~al.}(2011)\citenamefont
  {Hebeler}, \citenamefont {Bogner}, \citenamefont {Furnstahl}, \citenamefont
  {Nogga},\ and\ \citenamefont {Schwenk}}]{Hebeler2011}%
  \BibitemOpen
  \bibfield  {author} {\bibinfo {author} {\bibfnamefont {K.}~\bibnamefont
  {Hebeler}}, \bibinfo {author} {\bibfnamefont {S.~K.}\ \bibnamefont {Bogner}},
  \bibinfo {author} {\bibfnamefont {R.~J.}\ \bibnamefont {Furnstahl}}, \bibinfo
  {author} {\bibfnamefont {A.}~\bibnamefont {Nogga}}, \ and\ \bibinfo {author}
  {\bibfnamefont {A.}~\bibnamefont {Schwenk}},\ }\href {\doibase
  10.1103/PhysRevC.83.031301} {\bibfield  {journal} {\bibinfo  {journal} {Phys.
  Rev. C}\ }\textbf {\bibinfo {volume} {83}},\ \bibinfo {pages} {031301(R)}
  (\bibinfo {year} {2011})}\BibitemShut {NoStop}%
\bibitem [{\citenamefont {Simonis}\ \emph {et~al.}(2016)\citenamefont
  {Simonis}, \citenamefont {Hebeler}, \citenamefont {Holt}, \citenamefont
  {Men\'{e}ndez},\ and\ \citenamefont {Schwenk}}]{Simonis2016}%
  \BibitemOpen
  \bibfield  {author} {\bibinfo {author} {\bibfnamefont {J.}~\bibnamefont
  {Simonis}}, \bibinfo {author} {\bibfnamefont {K.}~\bibnamefont {Hebeler}},
  \bibinfo {author} {\bibfnamefont {J.~D.}\ \bibnamefont {Holt}}, \bibinfo
  {author} {\bibfnamefont {J.}~\bibnamefont {Men\'{e}ndez}}, \ and\ \bibinfo
  {author} {\bibfnamefont {A.}~\bibnamefont {Schwenk}},\ }\href {\doibase
  10.1103/PhysRevC.93.011302} {\bibfield  {journal} {\bibinfo  {journal} {Phys.
  Rev. C}\ }\textbf {\bibinfo {volume} {93}},\ \bibinfo {pages} {011302(R)}
  (\bibinfo {year} {2016})}\BibitemShut {NoStop}%
\bibitem [{\citenamefont {Simonis}\ \emph {et~al.}(2017)\citenamefont
  {Simonis}, \citenamefont {Stroberg}, \citenamefont {Hebeler}, \citenamefont
  {Holt},\ and\ \citenamefont {Schwenk}}]{Simonis2017}%
  \BibitemOpen
  \bibfield  {author} {\bibinfo {author} {\bibfnamefont {J.}~\bibnamefont
  {Simonis}}, \bibinfo {author} {\bibfnamefont {S.~R.}\ \bibnamefont
  {Stroberg}}, \bibinfo {author} {\bibfnamefont {K.}~\bibnamefont {Hebeler}},
  \bibinfo {author} {\bibfnamefont {J.~D.}\ \bibnamefont {Holt}}, \ and\
  \bibinfo {author} {\bibfnamefont {A.}~\bibnamefont {Schwenk}},\ }\href
  {\doibase 10.1103/PhysRevC.96.014303} {\bibfield  {journal} {\bibinfo
  {journal} {Phys. Rev. C}\ }\textbf {\bibinfo {volume} {96}},\ \bibinfo
  {pages} {014303} (\bibinfo {year} {2017})}\BibitemShut {NoStop}%
\bibitem [{\citenamefont {Holt}\ \emph {et~al.}(2019)\citenamefont {Holt},
  \citenamefont {Stroberg}, \citenamefont {Schwenk},\ and\ \citenamefont
  {Simonis}}]{Holt2019}%
  \BibitemOpen
  \bibfield  {author} {\bibinfo {author} {\bibfnamefont {J.~D.}\ \bibnamefont
  {Holt}}, \bibinfo {author} {\bibfnamefont {S.~R.}\ \bibnamefont {Stroberg}},
  \bibinfo {author} {\bibfnamefont {A.}~\bibnamefont {Schwenk}}, \ and\
  \bibinfo {author} {\bibfnamefont {J.}~\bibnamefont {Simonis}},\ }\href@noop
  {} {\  (\bibinfo {year} {2019})},\ \Eprint {http://arxiv.org/abs/1905.10475}
  {arXiv:1905.10475} \BibitemShut {NoStop}%
\bibitem [{\citenamefont {Stroberg}()}]{Stroberg2017b}%
  \BibitemOpen
  \bibfield  {author} {\bibinfo {author} {\bibfnamefont {S.~R.}\ \bibnamefont
  {Stroberg}},\ }\href {https://github.com/ragnarstroberg/imsrg} {\enquote
  {\bibinfo {title} {https://github.com/ragnarstroberg/imsrg},}\ }\BibitemShut
  {NoStop}%
\bibitem [{\citenamefont {Hergert}\ \emph {et~al.}(2016)\citenamefont
  {Hergert}, \citenamefont {Bogner}, \citenamefont {Morris}, \citenamefont
  {Schwenk},\ and\ \citenamefont {Tsukiyama}}]{Hergert2016}%
  \BibitemOpen
  \bibfield  {author} {\bibinfo {author} {\bibfnamefont {H.}~\bibnamefont
  {Hergert}}, \bibinfo {author} {\bibfnamefont {S.~K.}\ \bibnamefont {Bogner}},
  \bibinfo {author} {\bibfnamefont {T.~D.}\ \bibnamefont {Morris}}, \bibinfo
  {author} {\bibfnamefont {A.}~\bibnamefont {Schwenk}}, \ and\ \bibinfo
  {author} {\bibfnamefont {K.}~\bibnamefont {Tsukiyama}},\ }\href {\doibase
  10.1016/j.physrep.2015.12.007} {\bibfield  {journal} {\bibinfo  {journal}
  {Phys. Rept.}\ }\textbf {\bibinfo {volume} {621}},\ \bibinfo {pages} {165}
  (\bibinfo {year} {2016})}\BibitemShut {NoStop}%
\bibitem [{\citenamefont {Shimizu}\ \emph {et~al.}(2019)\citenamefont
  {Shimizu}, \citenamefont {Mizusaki}, \citenamefont {Utsuno},\ and\
  \citenamefont {Tsunoda}}]{Shimizu2019}%
  \BibitemOpen
  \bibfield  {author} {\bibinfo {author} {\bibfnamefont {N.}~\bibnamefont
  {Shimizu}}, \bibinfo {author} {\bibfnamefont {T.}~\bibnamefont {Mizusaki}},
  \bibinfo {author} {\bibfnamefont {Y.}~\bibnamefont {Utsuno}}, \ and\ \bibinfo
  {author} {\bibfnamefont {Y.}~\bibnamefont {Tsunoda}},\ }\href {\doibase
  10.1016/j.cpc.2019.06.011} {\bibfield  {journal} {\bibinfo  {journal}
  {Comput. Phys. Commun.}\ }\textbf {\bibinfo {volume} {244}},\ \bibinfo
  {pages} {372} (\bibinfo {year} {2019})}\BibitemShut {NoStop}%
\bibitem [{\citenamefont {Arthuis}\ \emph {et~al.}(2020)\citenamefont
  {Arthuis}, \citenamefont {Barbieri}, \citenamefont {Vorabbi},\ and\
  \citenamefont {Finelli}}]{Arthuis2020}%
  \BibitemOpen
  \bibfield  {author} {\bibinfo {author} {\bibfnamefont {P.}~\bibnamefont
  {Arthuis}}, \bibinfo {author} {\bibfnamefont {C.}~\bibnamefont {Barbieri}},
  \bibinfo {author} {\bibfnamefont {M.}~\bibnamefont {Vorabbi}}, \ and\
  \bibinfo {author} {\bibfnamefont {P.}~\bibnamefont {Finelli}},\ }\href@noop
  {} {\  (\bibinfo {year} {2020})},\ \Eprint {http://arxiv.org/abs/2002.02214}
  {arXiv:2002.02214 [nucl-th]} \BibitemShut {NoStop}%
\bibitem [{\citenamefont {Miyagi}\ \emph {et~al.}()\citenamefont {Miyagi} \emph
  {et~al.}}]{Miyagi}%
  \BibitemOpen
  \bibfield  {author} {\bibinfo {author} {\bibfnamefont {T.}~\bibnamefont
  {Miyagi}} \emph {et~al.},\ }\href@noop {} {\ }\bibinfo {note}
  {(unpublished)}\BibitemShut {NoStop}%
\bibitem [{\citenamefont {Miyagi}\ \emph {et~al.}(2020)\citenamefont {Miyagi},
  \citenamefont {Stroberg}, \citenamefont {Holt},\ and\ \citenamefont
  {Shimizu}}]{Miyagi2020}%
  \BibitemOpen
  \bibfield  {author} {\bibinfo {author} {\bibfnamefont {T.}~\bibnamefont
  {Miyagi}}, \bibinfo {author} {\bibfnamefont {S.}~\bibnamefont {Stroberg}},
  \bibinfo {author} {\bibfnamefont {J.}~\bibnamefont {Holt}}, \ and\ \bibinfo
  {author} {\bibfnamefont {N.}~\bibnamefont {Shimizu}},\ }\href@noop {} {\
  (\bibinfo {year} {2020})},\ \Eprint {http://arxiv.org/abs/2004.12969}
  {arXiv:2004.12969 [nucl-th]} \BibitemShut {NoStop}%
\bibitem [{\citenamefont {Taniuchi}\ \emph {et~al.}(2019)\citenamefont
  {Taniuchi} \emph {et~al.}}]{Taniuchi2019}%
  \BibitemOpen
  \bibfield  {author} {\bibinfo {author} {\bibfnamefont {R.}~\bibnamefont
  {Taniuchi}} \emph {et~al.},\ }\href {\doibase 10.1038/s41586-019-1155-x}
  {\bibfield  {journal} {\bibinfo  {journal} {Nature}\ }\textbf {\bibinfo
  {volume} {569}},\ \bibinfo {pages} {53} (\bibinfo {year} {2019})}\BibitemShut
  {NoStop}%
\end{thebibliography}%

\end{document}